\documentclass{iopjournal}
\usepackage{amsmath}
\usepackage{amssymb}
\usepackage{amsfonts}
\usepackage{array}
\usepackage{mathrsfs}
\usepackage{mathtools}
\usepackage{multirow}
\usepackage{siunitx}
\usepackage{subcaption}
\usepackage{booktabs}
\usepackage{ulem}
\usepackage{orcidlink}
\usepackage{soul}
\usepackage{aas_macros}

\begin{document}

\articletype{Paper}

\title{Impact on Inferred Neutron Star Equation of State due to Nonlinear Hydrodynamics, Background Spin, and Relativity}

\author{Joseph Bretz$^{1,*}$\orcidlink{0000-0001-5054-0070}, Hang Yu$^{1}$\orcidlink{0000-0002-6011-6190}}

\affil{$^1$Department of Physics, Montana State University, Bozeman, MT, USA}

\affil{$^*$Author to whom any correspondence should be addressed.}

\email{joseph.bretz@montana.edu}


\begin{abstract}
    Tidal interaction is a unique, detectable signature in gravitational wave signals from inspiraling binary neutron stars (BNSs), which can be used to constrain the neutron star (NS) equation of state (EoS).
    The tidal interaction is resonantly amplified as the orbital frequency approaches the NS fundamental mode (f-mode) frequency.
    It has been shown that the exclusion of tidal resonance in parameter estimation leads to a significant bias in the inferred NS tidal deformability and hence the NS EoS \cite{Pratten:22}.
    The strength and location of tidal resonance depend sensitively on the f-mode frequency, which is typically modeled using its universal relation with the tidal deformability that is derived for an isolated, non-spinning NS assuming only linear fluid perturbations.
    In a BNS inspiral, the f-mode frequency can be corrected by at least three known effects: nonlinear hydrodynamics, background spin, and relativity.
    We use Hamiltonian Monte Carlo simulations to estimate the systematic bias on tidal deformability when each frequency correction is ignored.
    Our study considers both loud, individual events and the stacking of a population of detections.
    Both scenarios are expected when the next-generation detectors are available with a sensitivity level increased by about an order of magnitude.
\end{abstract}

\section{Introduction}
\label{sec:intro}

Neutron stars (NSs) offer the opportunity to advance our understanding of fundamental physics because of the extreme conditions unique to NSs, which cannot be reproduced in terrestrial laboratories. In the core of a NS, densities exceed nuclear density and obey a currently unknown equation of state (EoS) for high-density, cold hadronic matter. One approach to probing the NS EoS is through the tidal interaction in a binary NS (BNS) inspiral. During a BNS inspiral, the two NSs are close enough that their gravitational fields induce a tidal deformation in each other which depends on the NS EoS. The BNS inspiral is driven by gravitational wave (GW) radiation, and as a result a tidal signature is imprinted on the observed GW signal. Tidal effects were included in the analysis of the first GW observation of a BNS merger, GW170817 \cite{GW170817, GW170817prop}, resulting in constraints on the supranuclear density EoS \cite{GW170817eos}. However, it has been shown that the exclusion of tidal resonance in these analyses of the tidal interaction can significantly bias the inferred EoS \cite{Pratten:22}. As more BNS merger events are observed and the next generation of GW detectors come online, such as the Cosmic Explorer \cite{Evans:17,Evans:21} and the Einstein Telescope \cite{ET:25,Punturo:10}, more sophisticated GW waveform models are needed to accurately constrain the NS EoS and advance our understanding of fundamental physics.

The dominant tidal effect in a BNS merger is the tidal field interacting with the fundamental mode (f-mode) of the NS, creating a large-scale deformation in the NS's shape. This effect was first studied in the adiabatic limit where the f-mode eigenfrequency is taken to be effectively infinite relative to the tidal driving frequency (the orbital frequency). In this limit the tidal deformation is characterized by a single coefficient, the tidal deformability, representing a proportionality constant between NS mass quadrupole and companion's tidal field. However, the tidal driving frequency does become comparable to the f-mode (eigen)frequency near merger. Models that keep the f-mode frequency as finite see an amplified tidal deformation of the NS due to tidal resonance. This finite-frequency (FF) correction to the tidal interaction significantly impacts the resulting GW waveform, and consequently the inferred NS EoS \cite{Pratten:22}. The strength and location of tidal resonance depend sensitively on the f-mode frequency, which is typically derived using only linear fluid perturbations. When modeling the GW waveform, the f-mode frequency is often reduced by its universal relation with the tidal deformability, derived for isolated, non-spinning NSs at the linear order in fluid perturbation. These simplifying assumptions overlook important corrections to the f-mode frequency, which are amplified by tidal resonance, leading to significant corrections in the GW waveform.

In a BNS inspiral, the f-mode frequency is shifted by at least three known effects: nonlinear hydrodynamics, NS background spin, and relativity. We refer to these corrections to the f-mode frequency as tidal resonance corrections (TRCs). The nonlinear hydrodynamic TRC we consider includes nonlinear tidal driving and coupling between eigenmodes, both of which effectively reduce the f-mode frequency \cite{Yu:23a}. For rotating NSs, the Doppler shift between inertial and co-rotating frames combines with the Coriolis force to effectively shift the f-mode frequency a fixed amount throughout the inspiral \cite{Kruger:20, Kruger:23, Steinhoff:21, Yu:24a, Yu:25a}. We also consider the relativistic effects of redshift and frame dragging and their impact on tidal resonance following the work of \cite{Steinhoff:16, Steinhoff:21}, which has been recently implemented in \cite{Haberland:25}. Even if a TRC does not shift the f-mode frequency to achieve resonance within the frequency band of a BNS inspiral, their impact on tidal resonance can still lead to significant corrections to the GW waveform.

In this work, we study the three types of TRCs described above and their impact on estimating NS parameters, with a focus on tidal deformability. To estimate detectability, we calculate the threshold signal-to-noise ratio (SNR) needed to discern between two waveforms: one waveform with the TRC and one waveform without. We perform this calculation for each type of TRC individually and for three different GW detectors' sensitivity curves -- Cosmic Explorer, Einstein Telescope, and Advanced LIGO A\#\footnote{An upgrade to Advanced LIGO following the O5 run.}. To estimate parameter bias due to excluding TRCs, we perform inference by fitting an injected waveform with a baseline model that includes linear tides but no TRCs. The injected waveform includes one type of TRC at a time. Inference is achieved through a Hamiltonian Monte Carlo (HMC) simulation. Our study considers both loud, individual events for all three types of TRCs, and a population of detections for the nonlinear hydrodynamics TRC.

The paper begins with an overview of the derivation of TRCs on the GW waveform, specifically the frequency-domain phase of the waveform in \S\ref{sec:waveform} , the phase shift due to tides in \S\ref{sec:tidal_phase_shift}, and the tidal excitation of f-modes in \S\ref{sec:hydrodynamical_response}. Next we cover the three types of TRCs that are the focus of this paper in \S\ref{sec:TRCs}. This is followed by integrating the TRCs into the GW waveform model and defining the baseline model in \S\ref{sec:tides_with_TRCs}. We then use our nonlinear-hydrodynamics TRC (derived in a Newtonian prescription) to compare with recent results derived using general relativity (GR) in the low-frequency regime in \S\ref{sec:low_frequency_expansion}, and outline some limitations of the effective Love number approach in \S\ref{sec:Eff_Love}. Next we define a canonical NS and dimensionless parameters in \S\ref{sec:canonical_NS} and take advantage of universal relations to reduce the number of free parameters in \S\ref{sec:UR}. The methods section ends with the waveform mismatch definitions for estimating detectability in \S\ref{sec:waveform_mismatch}, followed by our parameter estimation approach using HMC simulations in \S\ref{sec:parameter_estimation}. After methods, we present our results for the three TRCs in \S\ref{sec:results}, followed by conclusions and discussion in \S\ref{sec:conclusions}. Details for the full calculation of the tidal phase shift are presented in Appendix \ref{app:tidal_phase_shift_derivation}.

We adopt geometrical units, $G\!=\!c\!=\!1$. We denote the mass and radius of the primary NS as $M_\mathrm{A}$ and $R_\mathrm{A}$. The mass and radius of the companion NS is denoted as $M_\mathrm{B}$ and $R_\mathrm{B}$. We use a subscript ``$\mathrm{A}$'' for the primary NS parameters, and a subscript ``$\mathrm{B}$'' for the companion NS. To simplify notation we define $M_\mathrm{tot}\coloneq M_\mathrm{A} + M_\mathrm{B}$ as the total mass, $\mu\coloneq M_\mathrm{A} M_\mathrm{B}/M_\mathrm{tot}$ as the reduced mass, and $\eta\coloneq \mu/M_\mathrm{tot}$ as the symmetric mass ratio. For our analysis we work with the chirp mass and mass ratio: $\mathcal{M} \coloneq (M_\mathrm{A}^3 M_\mathrm{B}^3/M_\mathrm{tot})^{1/5}$ and $q \coloneq M_\mathrm{A}/M_\mathrm{B}$. Note we always take the primary NS to be of lower mass than the companion NS, $M_\mathrm{A}<M_\mathrm{B}$, hence $0<q<1$. For simplicity, we ignore radial dependence and take all NS radii to be a constant, $R\coloneq R_\mathrm{A}=R_\mathrm{B}=11.7$ km. 

Part of our analysis focuses on the impact of NS background spin. We denote the magnitude of the spin angular velocity of the primary NS as $\Omega_\mathrm{A}$. We restrict the scope of this work to background spin in only the primary NS, hence $\Omega_\mathrm{B}=0$. We take the NS spin angular velocity to be either aligned or anti-aligned with orbital angular velocity. Consequently, only the magnitude is needed, where $\Omega_\mathrm{A}>0$ signifies the NS spin and orbital angular velocities are aligned and $\Omega_\mathrm{A}<0$ signifies anti-alignment. We also use the dimensionless spin parameters, $\chi_\mathrm{A} \coloneq I_\mathrm{A}\,\Omega_\mathrm{A}\,/M_\mathrm{A}^2 = \bar{I}_\mathrm{A}\,\Omega_\mathrm{A}\,M_\mathrm{A}$, where $I_\mathrm{A}$ is the moment of inertia of the primary NS, and $\bar{I}_\mathrm{A}\coloneq I_\mathrm{A}/ \,M_\mathrm{A}^3$ is the dimensionless moment of inertia. We combine the dimensionless spin parameters to define an antisymmetric $\chi_- \coloneq \frac{1}{2}(\chi_\mathrm{A} - \chi_\mathrm{B})$ spin parameter and a mass-weighted, symmetric $\chi_+ \coloneq (q\,\chi_\mathrm{A}+\chi_\mathrm{B})/(1+q)$ spin parameter which is also commonly known as the effective spin.

For the orbit of the two NSs, we denote the magnitude of the orbital angular velocity as $\omega$ and the direction is taken to be fixed. We will often describe the orbital frequency in terms of the GW frequency due to quadrupolar GW radiation, which is twice the frequency of the orbit, $f_\mathrm{gw}=\omega/\pi$.

In our analysis, various simplification assumptions will be made (e.g., treating $R$ as a constant for all NSs). This means we only aim to provide a proof-of-principle study on the impacts of TRCs; a more careful treatment is deferred to the future. Indeed, no waveform model is currently available that includes all TRCs to high fidelity (e.g., \cite{Yu:23a} considered nonlinear hydrodynamics only in the Newtonian limit, while the nonlinear analyses by \cite{Pani:25} and \cite{Pitre:25} are restricted to the static or low-frequency limit; none of them considers spin effects.)

\section{Methods}

\subsection{Waveform}
\label{sec:waveform}

We focus our analysis on the GW signal from inspiraling BNSs , truncated at the NSs coming into contact and coalescing at merger. We refer to the inspiral of a BNS system as an event. We assume a quasi-circular orbit.
GWs are detected via the strain $h(t)$ measured at a detector. It is convenient to work with the Fourier transform of $h(t)$ in the frequency domain,
\begin{equation}
    \tilde{h}(f_{\mathrm{gw}}) \coloneq \int_{-\infty}^{\infty} e^{2 \pi i f_{\mathrm{gw}} t} h(t) d t .
\end{equation}
Here, $\tilde{h}(f_{\mathrm{gw}})$ is the frequency-dependent Fourier transform of the time-domain strain, $h(t)$. Note the chosen sign convention of the complex exponential.

Following \cite{Cutler:94}, we take advantage of the rapidly changing phase of the strain and use the stationary phase approximation to simplify the Fourier transform of the strain waveform,
\begin{equation}
    \tilde{h}(f_{\mathrm{gw}}) = \frac{Q}{D}\mathcal{M}^{5/6}f_{\mathrm{gw}}^{-7/6} e^{i \Psi(f_{\mathrm{gw}})} .
\end{equation}
The source's orientation and the detector's antenna response are contained within $Q$. The distance from the detector to the BNS system is $D$. We will specify the SNR of the GW signal instead of specifying $Q$ and $D$. The chirp mass $\mathcal{M}$ and frequency are the only remaining factors in the amplitude of $\tilde{h}$. Since the chirp mass is highly constrained by the early inspiral section of the GW signal, we take the chirp mass to be known and focus our analysis on the remaining parameters.

We further ignore the tidal correction to the amplitude of $\tilde{h}$ and focus on the phase, which can be split into a few terms,
\begin{equation} \label{eq:waveform-phase}
    \Psi(f_{\mathrm{gw}}) = \Psi_{\mathrm{pp}}(f_{\mathrm{gw}}) + \Psi_{\mathrm{tide}}(f_{\mathrm{gw}}) + 2\pi f_{\mathrm{gw}} t_{\mathrm{c}} - \varphi_{\mathrm{c}} \,,
\end{equation}
where $t_{\mathrm{c}}$ and $\varphi_{\mathrm{c}}$ are the time and phase of coalescence. We use $\varphi$ to denote the GW waveform phase and avoid confusion with the orbital phase, $\phi$. $\Psi_{\mathrm{pp}}$ is the point-particle (PP) contribution to the waveform, and $\Psi_{\mathrm{tide}}$ contains all of the tidal contribution.

For $\Psi_{\mathrm{pp}}$, we use the TaylorF2 waveform from the LALSuite library implemented in python via the ripple package \cite{rippleGithub:24}. We set all tidal parameters in TaylorF2 to zero, because the tidal contributions are contained within $\Psi_{\mathrm{tide}}$. Thus, the intrinsic PP phase only depends on the ratio of the NS masses and dimensionless spin parameters: $\Psi_{\mathrm{pp}}(f_{\mathrm{gw}},q,\chi_\mathrm{A},\chi_\mathrm{B})$.

For $\Psi_\mathrm{tide}$, we follow the derivations in \cite{Yu:23a} and \cite{Yu:24a} to quantify the tidal contribution to the waveform phase. The tidal contribution includes the three TRCs that are the focus of our study: nonlinear hydrodynamics presented in \cite{Yu:23a}, NS background spin effects from \cite{Yu:24a}, and the relativistic effects -- redshift and frame dragging -- as derived in \cite{Steinhoff:16, Steinhoff:21}.

\subsection{Tidal Phase Shift}
\label{sec:tidal_phase_shift}

To calculate $\Psi_\mathrm{tide}$, we use an energy-balancing approach \cite{Flanagan:08, Yu:23a}. We start with an equilibrium configuration that does not include tidal effects and show how the energy and energy loss due to GW radiation for such a configuration can be used to calculate the GW waveform. We then perturbatively include tides to calculate the corrections to the equilibrium configuration, and subsequently derive the tidal contribution to the waveform. This section provides an overview of the energy-balancing approach with more details for a full calculation presented in Appendix \ref{app:tidal_phase_shift_derivation}.

\textit{Point-Particle Configuration}$\quad$
We consider two NSs in a quasi-circular orbit around their common center of mass. We take the orbit to be in the x-y plane with the orbital angular momentum aligning with the positive z-axis. The origin of the coordinate system is at the center of the primary NS. We define an inertial frame of reference where the coordinate system is fixed even though the NSs may be spinning. The symmetry of the NS is conducive for spherical coordinates, so we define $\boldsymbol{r}$ as the vector pointing from the center of the primary NS to the center of the companion NS, and the magnitude, $r$, is the separation distance between the center of each NS. Restricting the orbit to the x-y plane enforces $\theta=\pi/2$ for the entire orbit. The azimuthal location of the companion star relative to the primary star, $\phi$, is also the orbital phase and is related to the orbital angular frequency, $\dot{\phi}=\omega$, where a dot over a variable denotes a time derivative.

Under Newtonian gravity, the BNS orbit is governed by the equations of motion for $r$ and $\phi$, following Eqs. (25) and (26) in \cite{Yu:23a}, 
\begin{eqnarray} \label{eq:eom_r_ddot}
    \ddot{r}-r \dot{\phi}^{2}+\frac{M_{\mathrm{tot}}}{r^{2}}&=&g_{r} \,,\\
    \label{eq:eom_phi_ddot}
    r \ddot{\phi}+2 \dot{r} \dot{\phi}&=&g_{\phi} \,.
\end{eqnarray}
We denote the non-Keplerian radial and tangential accelerations by $g_r$ and $g_\phi$. 

For the PP configuration, we only include GW radiation, a dissipative effect where orbital energy is radiated out of the system. Since the GW radiation leaks the orbital energy on a time scale much longer than the orbital period, we model the orbit as a Keplerian orbit that quasi-statically evolves with the loss of energy. This allows us to relate the orbital energy to the separation between the NSs, $r$, and thus the GW frequency. 

Since the equilibrium configuration does not include tides, the only reservoir of energy driving the dynamics of the system is the orbital energy, which we estimate as the orbital energy for two PPs following a Keplerian orbit, hence $E_{\mathrm{orb}}=E_{\mathrm{pp}}=-M_{\rm A} M_{\rm B}/(2r)$.

The orbital energy dissipates as GWs propagate outward, carrying energy out of the system. Consequently, orbital energy decreases, the separation between the NSs decreases, and the orbital frequency increases following Kepler's third law, $r^3=M_{\mathrm{tot}}/\omega^2$. The loss of orbital energy due to GW radiation is denoted as $\dot{E}$ and is equivalent to the GW radiation rate for a PP orbit, see Eq. (\ref{eq:omega_dot_pp}). 

If the orbital energy and GW radiation are known for all frequencies during the inspiral, then they can be combined to get the time and phase of coalescence,
\begin{equation}
    t\,=\int \frac{1}{\dot{E}} \frac{d E_{\mathrm{orb}}}{d f_\mathrm{gw}} d f_{\mathrm{gw}},
\end{equation}

\begin{equation}
    \varphi=2\pi \int \frac{f_{\mathrm{gw}}}{\dot{E}} \frac{d E_{\mathrm{orb}}}{d f_{\mathrm{gw}}} d f_{\mathrm{gw}}.
\end{equation}
The orbital frequency has been replaced with GW frequency using $\omega=\pi f_{\mathrm{gw}}$ in the relation between the differential time and phase of coalescence, $d\varphi_/d f_\mathrm{gw}=2\omega \,dt/d f_\mathrm{gw}$.

\textit{Tidal Corrections}$\quad$
The tidal contribution to the GW waveform can be thought of as frequency-dependent tidal corrections to the time and phase of coalescence, 
\begin{equation} \label{eq:psi_tide}
    \Psi_{\mathrm{tide}}=2\pi f_{\mathrm{gw}} \Delta t(f_\mathrm{gw})-\Delta \varphi[t(f_\mathrm{gw})] \,.
\end{equation}

To get the tidal contributions to the coalescence time and phase, we perturb the equilibrium energy and GW radiation,
\begin{equation} \label{eq:delta_tc}
\Delta t\, \simeq \int \frac{1}{\dot{E}_{\mathrm{pp}}}\left(\frac{d \Delta E}{d f_{\mathrm{gw}}}-\frac{d E_{\mathrm{orb}}}{d f_{\mathrm{gw}}} \frac{\Delta \dot{E}}{\dot{E}_{\mathrm{pp}}}\right) d f_{\mathrm{gw}}\,,
\end{equation}

\begin{equation} \label{eq:delta_phic}
\Delta \varphi \simeq 2\pi \int \frac{f_{\mathrm{gw}}}{\dot{E}_{\mathrm{pp}}}\left(\frac{d \Delta E}{d f_{\mathrm{gw}}}-\frac{d E_{\mathrm{orb}}}{d f_{\mathrm{gw}}} \frac{\Delta \dot{E}}{\dot{E}_{\mathrm{pp}}}\right) d f_{\mathrm{gw}} ,
\end{equation}
where the tidal corrections to the equilibrium energy and GW radiation are denoted $\Delta E$ and $\Delta\dot{E}$. 

We first describe the conservative effects due to the tidal interactions, arising from an interaction potential, $E_{\rm int}$, that couples the mode amplitude and tidal field. In the low-frequency limit, the interaction is primarily radial, leading to a modification to the orbital separation, $\Delta r$. This deviation from Kepler's third law further causes a correction to the orbital energy, $\Delta E_{\rm orb}$. As the tidal forcing frequency approaches the f-mode frequency, the tangential interaction (tidal torque) becomes important and dominates energy transfer from the orbit into the NS internal energy, denoted by $E_{\rm mode}$. Note that both the interaction energy $E_{\rm int}$ and the correction to the orbit $\Delta E_{\rm orb}$ see one factor of Lorentzian amplification due to mode resonance, whereas the internal NS energy $E_{\rm mode}$ sees the Lorentzian squared as it is proportional to the mode amplitude squared. Therefore, $E_{\rm mode}$ dominates near resonance. Collectively, the conservative perturbation to the system's equilibrium energy can be written as
\begin{equation} \label{eq:Delta_E}
\Delta E = \Delta E_{\rm orb} + E_{\rm int} + E_{\rm mode}\,.
\end{equation}

Next we describe the dissipative effects due to the tidal corrections to the GW radiation. The GW radiation is enhanced due to two main effects: the coupling of the tidally induced NS quadrupole with the orbital quadrupole, $\Delta \dot{E}_{\mathrm{ns-orb}}$; and the correction to the orbital quadrupole due to the orbital separation correction, $\Delta \dot{E}_{r-\omega}$. We ignore smaller effects on the GW radiation, such as the self-coupling between both NS's quadrupoles. In total, the tidal corrections to the GW radiation are
\begin{equation} \label{eq:Delta_E_dot}
\Delta \dot{E}=\Delta \dot{E}_{\mathrm{ns-orb}}+\Delta \dot{E}_{r-\omega}\,.
\end{equation}

\cite{Yu:23a} quantify the orbital separation correction to the Keplerian orbit as $O[\Delta r/r] \propto (R/r)^5$.  They also show that the leading-order non-linear tidal correction (terms proportional to fluid displacement or mode amplitude squared in the equation of motion or cubed in the energy) enters at $(R/r)^{8}$, thus only one order in $\Delta r/r$ is needed. Terms $\propto (\Delta r/r)^2\propto (R/r)^{10}$ are neglected in this analysis, though \cite{Yu:24a} suggested that they are significant when the f-mode is close to resonance. 

\subsection{Hydrodynamical Response}
\label{sec:hydrodynamical_response}

\begin{figure}
    \centering
    \includegraphics[width=0.5\linewidth]{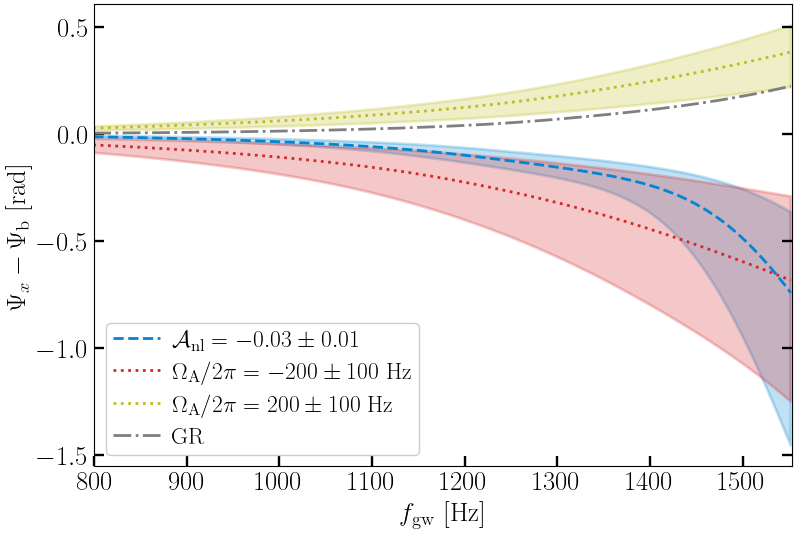}
    \caption{
        The impact of TRCs on the frequency-domain GW waveform phase. The difference between $\Psi_x$ -- the waveform phase with the TRC denoted in the legend -- and $\Psi_\mathrm{b}$ -- the baseline waveform phase with FF corrections but no TRC -- is plotted in radians against GW frequency in Hz up to approximate merger. 
        The nonlinear hydrodynamics TRC (blue dashed) reduces the f-mode frequency more strongly at higher orbital frequency, resulting in a reduction in waveform phase that sharply grows at high $f_\mathrm{gw}$. 
        The anti-aligned background spin TRC in the primary NS (red dotted) reduces the f-mode frequency by a fixed amount, resulting in a more gradual reduction in waveform phase. 
        The aligned background spin TRC (olive dotted) does the opposite, increasing the waveform phase.
        The GR TRC (gray dot-dashed) increases the f-mode frequency, resulting in less resonance and a longer inspiral relative to the baseline.
        }
    \label{fig:f-mode effects}
\end{figure}

To calculate the hydrodynamical response of the NS, we now work in the co-rotating frame. While still centered on the primary NS, the co-rotating frame rotates with the NS at its constant spin frequency, $\Omega_\mathrm{A}$. We calculate the tidal amplitude in the primary NS by starting with a Lagrangian displacement, $\boldsymbol{\xi}$, to linear order in displacement following \cite{Schenk:02}.
We then proceed with the eigenmode decomposition of perturbations on the primary NS,\footnote{While such a modal approach is rooted in Newtonian hydrodynamics and has limitations in full relativity \cite{Pitre:24, Pitre:25, Andersson:25, Miao:25, Pani:25}, it is the only methodology available at the moment that allows us to probe high-frequency dynamics including resonances. } where the modes are denoted by the subscript $a$,
\begin{equation}
    \left[
    \begin{array}{l}
        \boldsymbol{\xi} \\
        \dot{\boldsymbol{\xi}}
    \end{array}
    \right]=\sum_{a} q_{a}\left[
    \begin{array}{c}
        \boldsymbol{\xi}_{a} \\
        -i \omega_{a\mathrm{S}} \boldsymbol{\xi}_{a}
    \end{array}
    \right].
\label{eq:mode_decomposition}
\end{equation}
Here $\boldsymbol{\xi}_{a}$ are the spatial eigenfunctions of the modes, $q_a$ are the modes' amplitudes, and $\omega_{a\mathrm{S}}$ are the eigenfrequencies in the co-rotating frame. If the NS background spin is smaller than the non-spinning f-mode frequency, then to linear order in spin the eigenfrequencies in the co-rotating frame, $\omega_{a\mathrm{S}}$, are related to the non-spinning eigenfrequencies, $\omega_{a0}$, by Eq. (\ref{eq:omega-co-rotating}).

Each mode has four quantum numbers, $a=\{n_a,l_a,m_a, s_a\}$, which correspond to nodes in $\boldsymbol{\xi}_a$ for the radial, polar, and azimuthal directions, $\{r,\theta,\phi\}$, as well as the sign of the eigenfrequency; the phase space expansion includes modes with both positive and negative frequencies. Our analysis focuses on the f-modes with $n_a=0$ which strongly dominate the tidal coupling in NSs. We also focus on the modes with the strongest coupling during the inspiral, the quadrupolar modes, $l_a=2$. This leaves three positive-frequency modes, $m=\{2,0,-2\}$, which are prograde, stationary, and retrograde respectively, as well as their complex conjugates obtained by simultaneously flipping the signs of $m_a$ and $\omega_{a\mathrm{S}}$. We ignore all other modes.

Using the orthogonality condition $\omega_{a0} (\omega_{a0} + \omega_{b0}) \int d^3x \rho \,\boldsymbol{\xi}_a^\ast \cdot \boldsymbol{\xi}_b=E_\mathrm{A}\delta_{ab}$, where $E_\mathrm{A}\coloneq M_\mathrm{A}^2/R$, \cite{Yu:23a} simplify the governing equation of the modes to
\begin{equation} \label{eq:mode_gov_eq}
    \dot{q}_{a}+i \omega_{a\mathrm{S}}\, q_{a}=i \omega_{a 0} U_{a} + \sum_b i\omega_{a0} U_{ab}^\ast \,q_b^* + \sum_{bc} i\omega_{a0} \kappa_{abc}\,q_b^*q_c^*\,.
\end{equation}
$U_{a}$ and $U_{ab}$ are the linear and nonlinear tidal overlaps with the eigenmodes, shown below. $\kappa_{abc}$ are the coupling coefficients between different eigenmode combinations, {$a,b,c$} \cite{Weinberg:12}. The left side describes a harmonic oscillator driven by the terms on the right side -- the couplings with the tidal potential and the three-mode couplings. These terms cover all the non-linear corrections to the mode amplitude to the $(\xi/R)^2 \sim (R/r)^6$ order in the equations of motion.

The linear and nonlinear tidal overlap terms include the geometry and time-dependence of the orbit. We express the linear and nonlinear tidal overlaps as,
\begin{equation} \label{eq:Ua}
    U_{a}=\frac{M_\mathrm{B}}{M_\mathrm{A}} W_{l_a m_a} I_{a,lm}\left(\frac{R}{r}\right)^{l_a+1} e^{-i m_a (\phi-\Omega_\mathrm{A} t)}=V_{a} e^{-i m_a (\phi-\Omega_\mathrm{A} t)}\,,
\end{equation}
\begin{equation} \label{eq:U_ablm}
    U_{ab,lm}=\frac{M_\mathrm{B}}{M_\mathrm{A}} W_{l m} J_{ab,lm}\left(\frac{R}{r}\right)^{l+1} e^{-i m (\phi-\Omega_\mathrm{A} t)}=V_{ab,lm} e^{-i m (\phi-\Omega_\mathrm{A} t)}\,,
\end{equation}
where we have pulled out the time dependence of the orbit to define $V_a$ and $V_{ab,lm}$. Note the quantum numbers with subscript ``$a$'' denote the \textit{primary} NS modes, while the quantum numbers $l$ and $m$ with no subscript denote the spherical decomposition of the tidal potential of the \textit{companion} NS. The $W_{lm}\coloneq4 \pi Y_{lm}(\pi/2, 0)/(2l+1)$ are fixed coefficients.
The integration over the eigenfunctions is contained within the $I_{a,lm}$ and $J_{ab,lm}$ terms,
\begin{equation} \label{eq:Ia}
    I_{a,lm}=\frac{\delta_{l_al}\delta_{m_am}}{M_\mathrm{A} R^{l}} \int \mathrm{~d}^{3} x \,\rho\, \boldsymbol{\xi}_{a}^{*} \cdot \nabla\left(r^l Y_{lm}\right) \,,
\end{equation}
\begin{equation} \label{eq:J_ablm}
    J_{ab,lm}=\frac{1}{M_\mathrm{A} R^{l}} \int \mathrm{~d}^{3} x \,\rho\, \boldsymbol{\xi}_{a} \cdot\left(\boldsymbol{\xi}_{b} \cdot \nabla\right) \nabla\left(r^{l} Y_{l m}\right) \,.
\end{equation}
The tidal potential from the companion star is $r^l Y_{lm}$, and $\rho$ is the stellar density. 

To find the solutions to this driven harmonic oscillator, we follow \cite{Yu:24a} and decompose the time-dependent mode amplitudes into two terms representing the equilibrium tide and the dynamical tide,
\begin{equation} \label{eq:eq-dyn tide}
    q_a = b_{a}^{(\mathrm{eq})} e^{-i m_{a}(\phi-\Omega_\mathrm{A} t)}+c_{a}^{(\mathrm{dyn})} e^{-i\left(\omega_{a\mathrm{S}} t-\psi_\mathrm{r}\right)} \,.
\end{equation}
The equilibrium tide varies with the driving force, the tidal field of the orbiting companion NS, encapsulated in the time dependence of the orbital phase, $\phi$. The orbital phase is shifted by the NS background spin, $\Omega_\mathrm{A}$.
The dynamical tide oscillates at the f-mode eigenfrequency, $\omega_{a\mathrm{S}}$.\footnote{Note our convention follows \cite{Yu:24a} and differs from \cite{Hinderer:16}. FF corrections (\textit{i.e.}, the Lorentzian) is called the dynamical tide in \cite{Hinderer:16}, but it is part of the equilibrium tide in our convention, as it is still phase-coherent with the orbit and represents the particular solution of a forced oscillator. The dynamical tide, in our case, specifically refers to the resonantly excited component and represents the homogeneous solution.
}
Alternative ways of obtaining mode solutions valid in the presence of resonance can be found in, e.g., \cite{Pnigouras:25}.

The phase of the dynamical tide is shifted by $\psi_\mathrm{r}$, which is chosen such that the phase goes to 0 at resonance for ease of numerical calculation. Resonance refers to when the driving frequency equals the NS's f-mode frequency, including corrections due to nonlinear hydrodynamics, background spin, or GR. The resonance term in the phase of the dynamical tide is $\psi_\mathrm{r} \coloneq (\omega_{a\mathrm{S}} + m_a \Omega_\mathrm{A})t_\mathrm{r} -m_a\phi_\mathrm{r}$, where $t_{\rm r}$ and $\phi_{\rm r}$ are respectively the time and orbital phase when resonance occurs.

What's left is to solve for the equilibrium and dynamical tide amplitudes, $b_{a}^{(\mathrm{eq})}$ and  $c_{a}^{(\mathrm{dyn})}$, which we present following our implementation of the three TRCs -- nonlinear hydrodynamics, background spin, and the relativistic effects redshift and frame dragging.

\subsection{Tidal Resonance Corrections}
\label{sec:TRCs}

\subsubsection{Nonlinear Hydrodynamics}
\label{sec:nonlinear_tides} 

\cite{Yu:23a} show that nonlinear tidal corrections due to coupling between mode amplitude and the tidal potential lead to an effective shift in the f-mode frequency which depends on the orbital frequency,
\begin{equation} \label{eq:Delta_wnl_J_k}
    \frac{\Delta \omega_{\mathrm{nl}}}{\omega_{a0}}=\sqrt{\frac{\pi}{5}}\left(J_{2}+4 \kappa_{2} I_{a}\right) \frac{R^{3}}{M_\mathrm{tot}} \frac{M_\mathrm{B}}{M_\mathrm{A}}\,\omega^{2} \,,
\end{equation}
where the coupling coefficients, $J_2=-0.21$ and $\kappa_2=-0.45$, are for the Newtonian $\Gamma=2$ polytrope given in table 1 of \cite{Yu:23a}. $J_2$ encapsulates the nonlinear tidal overlap, see Eq. (\ref{eq:J_ablm}); and $\kappa_2$ represents the coupling between three eigenmodes of the star with azimuthal quantum numbers $(2, 0, -2)$; both calculated using equations in the appendix A of \cite{Weinberg:12}; they also agree well with the affine model's prediction of \cite{Yu:26a}. The tidal overlap integral in the second term on the right, given by Eq. (\ref{eq:Ia}) is related to the static tidal Love number, $k_{2} \simeq \frac{4 \pi}{5} I_{a}^{2}$.

Note $\Delta\omega_\mathrm{nl}/\omega_{a0}<0$, resulting in a lower effective f-mode frequency. The quadratic dependency on orbital frequency implies the downward shift in f-mode frequency grows as the BNS inspiral evolves. See appendix A in \cite{Yu:23a} for an explanation of the physical origin of the f-mode shift due to nonlinear hydrodynamics.

There are also nonlinear hydrodynamics corrections to the tidal drive, denoted by $\Delta V$ in \cite{Yu:23a}. However these corrections do not undergo resonant amplification and become subdominant to the f-mode frequency shift corrections near merger; see figure 3 in \cite{Yu:23a}. Nonlinear corrections beyond $\Delta \omega_\mathrm{nl}$ are hence ignored in our proof-of-principle analysis for simplicity. Note the $\Delta V$ corrections add to the $\Delta\omega_\mathrm{nl}$ corrections, so dropping them underestimates the total impact of nonlinear hydrodynamics on the waveform phase; see Fig. \ref{fig:eff_love} for a visual comparison between our model with (orange) and without (blue) the tidal drive corrections.

To parameterize the nonlinear hydrodynamics TRC, we define a dimensionless magnitude at a specified reference frequency,
\begin{equation} \label{eq:Delta_wnl_Anl_wref}
    \frac{\Delta \omega_\mathrm{nl}}{\omega_{a0}} \coloneq \mathcal{A}_\mathrm{nl} \left(\frac{\omega}{\omega_\mathrm{ref}}\right)^2 \,.
\end{equation}
We choose a reference GW frequency of 1 kHz, $\omega_\mathrm{ref}=\pi(1000\mathrm{\;Hz})$, which results in a magnitude of $\mathcal{A}_\mathrm{nl}\simeq-0.03$, following figure B1 in \cite{Yu:23a}.

\subsubsection{Neutron Star Background Spin}

NS background spin modifies the f-mode frequency in two ways -- Doppler shift and the Coriolis force.
The Doppler effect enters as a shift of the companion's angular position in the co-rotating frame where the eigenmodes are defined. In this frame, a mode is forced with a driving frequency of $m_a(\dot{\phi} - \Omega_{\rm A})$.
The Coriolis force directly modifies the internal restoring force of a mode.
Assuming the NS background spin is small compared to the non-spinning f-mode frequency, then to linear order the eigenfrequency in the co-rotating frame, $\omega_{a\rm{S}}$, is related to the non-spinning eigenfrequency, $\omega_{a0}$, as \cite{Yu:24a}
\begin{equation} \label{eq:omega-co-rotating}
    \omega_{a\mathrm{S}} \simeq \omega_{a 0} -\frac{m_{a}}{l_{a}} \Omega_\mathrm{A}\,.
\end{equation}
By plugging Eq. (\ref{eq:omega-co-rotating}) into Eq. (\ref{eq:mode_gov_eq}), one can see the Doppler and Coriolis effects counteract each other, with the Doppler shift being dominant.
The two effects also combine in Eq. (\ref{eq:omega_r^0_second}) which represents the shifted tidal resonance, $\omega_\mathrm{r}^{(0)}$, due to the background spin TRC. 
Both the Doppler and Coriolis effects are proportional to the NS spin (for small spin) and combine to offset the resonance from the non-spinning f-mode frequency, $\omega_{a0}$.
Thus a NS with spin anti-aligned with the orbital angular momentum will effectively reduce the f-mode frequency, and amplify the resonance as the BNS approaches merger.

\subsubsection{Redshift and Frame Dragging}

We include two relativistic corrections to the f-mode frequency, redshift and frame dragging. First is redshift which we implement via the expansion in orbital frequency from \cite{Steinhoff:21},
\begin{equation} \label{eq:redshift}
    z\simeq  1+\frac{x}{2}X_\mathrm{B}\left(X_\mathrm{A} - 3\right) +\frac{x^{2}}{24} X_\mathrm{B}\left(5 X_\mathrm{A}^2 X_\mathrm{B} - 9 X_\mathrm{A} - 6 X_\mathrm{A} X_\mathrm{B} - 27\right),
\end{equation}
where $x\coloneq(M_\mathrm{tot}\omega)^{2/3}$ is a scaled, dimensionless orbital frequency, and $X_i\coloneq M_i/M_\mathrm{tot}$ are the NS mass fractions. Redshift effectively reduces the f-mode frequency, since $X_\mathrm{A}<1$.

We also use the orbital-frequency dependent frame dragging from \cite{Steinhoff:21},
\begin{equation} \label{eq:frame_dragging}
    \frac{\Omega_{\mathrm{FD}}}{\omega}\simeq \frac{x}{2} X_\mathrm{B}\left(X_\mathrm{A}+3\right)-\frac{x^{2}}{24} X_\mathrm{B}\left(X_\mathrm{A}^{2}X_\mathrm{B}-45 X_\mathrm{A}+30 X_\mathrm{A} X_\mathrm{B}-27\right)\,.
\end{equation}
Note we do not include the spin-spin interaction which enters at $\mathcal{O}(x^{3/2})$, as we only consider one type of TRC at a time.
In the case of no GR, we set $z\rightarrow 1$ and $\Omega_{\mathrm{FD}}\rightarrow 0$. Similar to the treatment of the nonlinear tide, we ignore corrections to the tidal drive. 

\subsection{Tides With Resonance Corrections}
\label{sec:tides_with_TRCs}

\subsubsection{Equilibrium Tides}

Following \cite{Yu:23a, Yu:24a, Yu:25a} and equation (5.3) from \cite{Steinhoff:21}, and using the equilibrium time dependence in Eq. (\ref{eq:eq-dyn tide}), the governing equation for f-modes including TRCs becomes,
\begin{equation} \label{eq:equil_mode_gov_eq}
    \dot{b}_{a}^{(\mathrm{eq})}+i \left[z\,\omega_{a\mathrm{S}}-m_a(\omega-z\,\Omega_\mathrm{A}) + m_a\Omega_\mathrm{FD} + \mathcal{A}_\mathrm{nl} \left(\frac{\omega}{\omega_\mathrm{ref}}\right)^2\omega_{a0}\right]\, b_{a}^{(\mathrm{eq})} = i z\,\omega_{a 0} V_{a}\,,
\end{equation}
where we've pulled out the time dependence of the linear tidal driving potential, $U_a=V_a e^{-im_a(\phi-\Omega_\mathrm{A} t)}$, which matches the time dependence of the equilibrium tide. Note the nonlinear tidal driving and three-mode coupling terms are contained within the parameterized nonlinear hydrodynamics TRC term scaled by $\mathcal{A}_{\mathrm{nl}}$. We remind the readers that corrections to $V_a$ are dropped in this proof-of-principle study for simplicity.

To solve for $b_a^{(\mathrm{eq})}$ we take advantage of it varying slowly with time prior to resonance and drop the $\dot{b}_a^{(\mathrm{eq})}$ term and calculate the zeroth order solution to the equilibrium tide amplitude,
\begin{equation} \label{eq:b_a0}
    b_{a}^{(0)} \simeq \frac{z\,\omega_{a 0} }{\Delta_{a}} V_{a} \, .
\end{equation}
We simplify the notation by defining a detuning factor,
\begin{equation} \label{eq:Delta_a}
    \Delta_{a} \coloneq z\left(\omega_{a\mathrm{S}}+m_{a} \Omega_\mathrm{A}\right)-m_{a} \omega + m_{a} \Omega_{\text{FD}} + \mathcal{A}_{\mathrm{nl}} \left(\frac{\omega}{\omega_{\text{ref}}}\right)^{2} \omega_{a0} \,.
\end{equation}
From here, our calculation follows \cite{Yu:24a}, where the first order correction to the equilibrium tide amplitude is found by estimating the full amplitude as $b_{a}^{(\mathrm{eq})} =  b_{a}^{(0)} + b_{a}^{(1)} + \cdots \simeq b_{a}^{(0)}/(1-b_{a}^{(1)}/b_{a}^{(0)})$, where we resum the series under the inspiration of \cite{Lai:94c}.
We get the same result for $b_a^{(1)}$ as \cite{Yu:24a}, because the time dependence of the additional effects included here is doubly small and dropped. 
We now resum the equilibrium tide amplitude and get a result very similar to equation (18) in \cite{Yu:24a} but with a different dephasing factor, $\Delta_a$, and a redshift factor:
\begin{equation} \label{eq:eq_tide}
    b_{a}^{(\mathrm{eq})} \simeq \frac{z \Delta_{a} \omega_{a 0} V_{a}}{\Delta_{a}^{2}-i\left[m_{a} \dot{\omega}-\Delta_{a}\left(l_{a}+1\right) \frac{\dot{r}}{r}\right]} \,.
\end{equation}
Resonance occurs when the real part of the denominator goes to zero, hence $\Delta_a\rightarrow0$. The imaginary component of the denominator acts as effective damping, preventing the amplitude from diverging at resonance without needing any fluid dissipation. This damping is due to the finite amount of time the system is at resonance, as the evolving orbit will pass through resonance over a time scale encoded by the time derivatives, $\dot{\omega}$ and $\dot{r}$. Note at resonance, $b_{a}^{(\mathrm{eq})}=0$.

\subsubsection{Dynamical Tides}

To calculate the dynamical tide amplitude, we combine Eqs. (\ref{eq:eq_tide}), (\ref{eq:eq-dyn tide}) and (\ref{eq:mode_gov_eq}) to get an equation of motion for the dynamical tide, see eqs. (19, 20) in \cite{Yu:24a}. We follow the same steps as \cite{Yu:24a} to arrive at the solution for the dynamical tide which nearly matches their result in their eq. (23),
\begin{equation} \label{eq:dyn_tide}
    c_{a}^{(\mathrm{dyn})} \simeq \frac{z \,\omega_{a 0} V_{a,2m,\mathrm{r}}}{\sqrt{\dot{\omega}_{\mathrm{pp},\mathrm{r}}}} F(u) \,,
\end{equation}
where $F(u)$ is a numerical integral over a scaled time factor, $u$, see Eqs. (\ref{eq:Fu}) and (\ref{eq:u}). Since tidal perturbations are kept to first order, the PP orbit can be used for $\dot{\omega}$. The subscript ``$\rm{r}$'' denotes evaluation at resonance. See Appendix \ref{app:tidal_phase_shift_derivation} for the full calculation.

What remains is calculating the resonance frequency which we estimate by setting $\Delta_a=0$ and taking the f-mode frequency corrections as small. For $m_{a}=2$, the zeroth order resonance frequency encapsulates the NS background spin,
\begin{eqnarray} \label{eq:omega_r^0}
    \omega_\mathrm{r}^{(0)} &=& \left(\frac{\omega_{a\mathrm{S}}}{2} + \Omega_\mathrm{A}\right) \\ \label{eq:omega_r^0_second}
    &\simeq& \frac{\omega_{a0}}{2}+
    \frac{\Omega_\mathrm{A}}{2} \,.
\end{eqnarray}
The second line comes from replacing the co-rotating eigenfrequency with Eq. (\ref{eq:omega-co-rotating}) for $l=m=2$. When not including the background spin TRC, we set $\Omega_\mathrm{A}\rightarrow 0$, and the zeroth order resonance frequency reduces to $\omega_{a0}/2$. From Eq. (\ref{eq:omega_r^0_second}), one can see how the tidal resonance frequency is directly shifted by the background spin frequency.

Next, we substitute the zeroth order solution into Eq. (\ref{eq:Delta_a}) and keep to first order to get the corrected resonance due to GR and nonlinear hydrodynamics,
\begin{equation}\label{eq:omega_r^1}
    \omega_{\mathrm{r}}^{(1)} = \left(1+ \eta \left(M_{\mathrm{tot}}\,\omega_\mathrm{r}^{(0)}\right)^{2/3}gr\right)\left(\omega_\mathrm{r}^{(0)} + \mathcal{A}_\mathrm{nl} \left(\frac{\omega_\mathrm{r}^{(0)}}{\omega_{\mathrm{ref}}}\right)^{2} \frac{\omega_{a0}}{2}\right) \,.
\end{equation}
When including the GR TRC in the corrected tidal resonance frequency, we set $gr\rightarrow 1$; otherwise $gr\rightarrow 0$. Likewise, when not including the nonlinear hydrodynamics TRC, we set $\mathcal{A}_\mathrm{nl}\rightarrow 0$. Eq. (\ref{eq:omega_r^1}) shows how the GR TRC increases the tidal resonance frequency, while the nonlinear hydrodynamics TRC decreases it ($\mathcal{A}_\mathrm{nl}<0$). and background spin shifts the resonance by the spin frequency.

\subsubsection{Tidal Corrections to Energy and GW Radiation}

\begin{figure}
    \centering
    \includegraphics[width=0.75\linewidth]{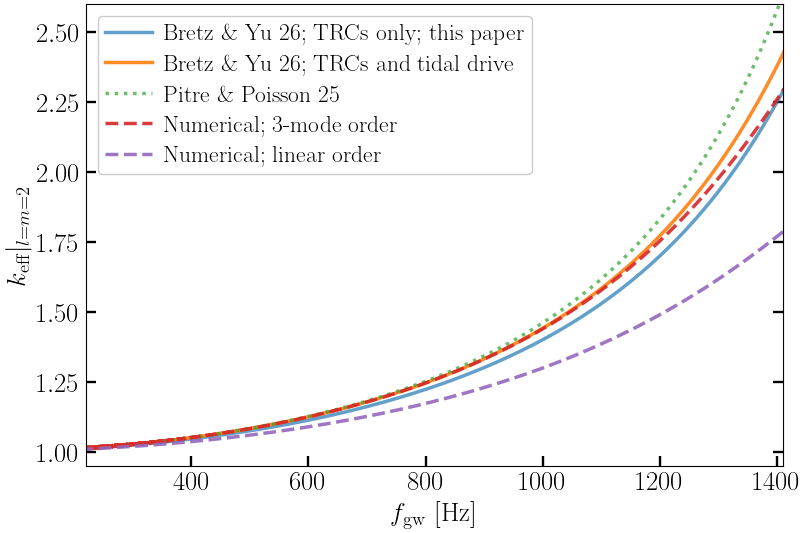}
    \caption{
    Comparison of the $l=m=2$ effective Love numbers for various nonlinear hydrodynamics models. 
    The computation assumes a Newtonian ($M_{\rm A}/R_{\rm A}=0$), $\Gamma=1+1/n=2$ polytrope. 
    (Blue) The model used in this paper with only TRCs, (orange) same as blue but with tidal drive corrections, (green-dotted) the resummed effective Love number from \cite{Pitre:25}, (red-dashed) the numerical model including nonlinear hydrodynamics to the 3-mode order, (purple-dashed) the numerical model without nonlinear corrections.
    }
    \label{fig:eff_love}
\end{figure}

With solutions to the equilibrium and dynamical tide amplitudes, the tidal phase shift to the GW waveform, Eqs. (\ref{eq:psi_tide}-\ref{eq:Delta_E_dot}), can now be calculated using the orbital energy and GW radiation with tidal corrections. While we include various corrections when deriving the mode amplitudes, we still use the linear, Newtonian form for the back-reaction. That is, we ignore post-Newtonian and nonlinear interaction potentials when computing the equilibrium configurations of the system. While formally this is inaccurate, previous results from figs. 4-6 of \cite{Yu:23a} suggest that numerically, the approximation might be fine. We point out that our study here is a proof-of-principle study that scopes out the qualitative impacts of various effects. A more accurate analysis should be done using more accurate waveform models like \cite{Yu:25a} and its nonlinear extensions (work in prep.). Yet those models are expensive to generate and hence not used here.

To assess the accuracy of our model and simplifying assumptions, we calculate the numerical solution to the f-mode amplitude truncated at the 3-mode order, Eq. (\ref{eq:mode_gov_eq}), by evolving it under a PP orbit driven by GW radiation, see Appendix \ref{app:eff_love_numerical}. The NS is assumed to be a $\Gamma=1+1/n=2$ polytrope. The resulting effective Love numbers (see later in Eq. \ref{eq:tidal_response}) are plotted in Figure \ref{fig:eff_love}, where our model with TRCs in blue is compared against the full model including tidal drive corrections in orange, the numerical model with nonlinear corrections in red dashed, and without in purple dashed. Fig. \ref{fig:eff_love} shows how our conservative choice to drop the tidal drive terms (blue-solid) slightly underestimates the effective Love number with tidal drive terms (orange), while capturing the significant correction from the numerical linear case (purple-dashed) to the numerical nonlinear 3-mode case (red-dashed).

An example of a subdominant nonlinear correction is the nonlinear driving term in the tidal interaction energy in eq. (27) of \cite{Yu:23a}. We thus drop the $U_{ab}$ terms to calculate the tidal interaction energy with the equilibrium tidal amplitude,
\begin{equation} \label{eq:Eint_lin}
    \frac{E_{\mathrm{int}}}{E_\mathrm{A}} \simeq - \sum_{a}^{m_a=m} V_a b_a^* \,.
\end{equation}
Here the summation is over the $a$ modes with both positive and negative frequencies.

We can also drop the subdominant nonlinear term in the tidal mode energy. The $\kappa_{abc}$ terms in eq. (43) of \cite{Yu:23a} can be dropped to give the tidal mode energy in terms of the tidal amplitude,
\begin{equation} \label{eq:Emode_lin}
    \frac{E_{\mathrm{mode}}}{E_{\mathrm{A}}} \simeq \frac{1}{2} \sum_{a} \frac{\omega_{a\mathrm{S}}+m_a\Omega_{\mathrm{A}}}{\omega_{a0}} \left| q_a \right|^2 \,.
\end{equation}
Converting the mode energy from the co-rotating frame to the inertial frame introduces the factor of $\Omega_\mathrm{A}$, see eq. (58) of \cite{Yu:24a}. Again the summation is over the $a$ modes for both positive and negative frequencies.

The energy and GW radiation corrections due to the modified Kepler's law are all proportional to the NS separation correction ($\Delta r$) given by eq. (39) in \cite{Yu:23a}. We drop the subdominant $J_{ab,lm}$ terms to get
\begin{equation} \label{eq:Delta-r}
    \frac{\Delta r}{r} \simeq \left( \frac{R}{r} \right)^{\!2}\,\sum_{m}^{\{2,0,-2\}} W_{2m} \sum_{a}^{m_a = m} I_{a,2m} \operatorname{Re}[b_a]\,,
\end{equation}
where $I_{a,2m}$ is the linear tidal overlap between tidal mode $a=\{l_a=2,m_a=m\}$ and tidal field moment $\{l=2,m\}$. The first summation is over the tidal field moments, and the second summation is over the tidal modes with equal azimuthal quantum numbers. The tidal correction to the orbital energy is given by eq. (48) in \cite{Yu:23a} as $\Delta E_\mathrm{orb}/E_\mathrm{orb}=-4 \Delta r/r$. The tidal correction to the GW radiation is given by eq. (59) in \cite{Yu:23a} as $\Delta \dot{E}_{r-\omega}/\dot{E}_\mathrm{pp}=4\Delta r/r$.

For the GW radiation correction due to NS-orbit coupling, we drop the $J_{ab,lm}$ terms in eq. (56) of \cite{Yu:23a}. The GW radiation correction for NS-orbit coupling reduces to
\begin{equation} \label{eq:Edot_ns-orb}
    \Delta \dot{E}_{\mathrm{ns}-\mathrm{orb}}=-\frac{8}{15} \eta \frac{M_\mathrm{A} R^{2}}{M_\mathrm{tot}^3} \left(M_\mathrm{tot}\omega\right)^{14/3} \sum_{m}^{\{2,-2\}} m^{6} W_{2m} \sum_a^{m_a=m} I_{a,2m} \operatorname{Re}\left[b_{a}\right] \,.
\end{equation}
where the first summation over the tidal field moments does not include the $m=0$ term due to the factor of $m^6$. The second summation is the same as the previous equation. $\eta$ is the symmetric mass ratio, defined at the end of \S\ref{sec:intro}. The justifications of above simplifications can be found from figs. 4-6 of \cite{Yu:23a}.

We combine the dominant tidal corrections to the equilibrium energy and GW radiation presented in this section via Eqs. (\ref{eq:Delta_E}, \ref{eq:Delta_E_dot}). From there the tidal contribution to the GW waveform phase is calculated using Eqs. (\ref{eq:psi_tide}-\ref{eq:delta_phic}).

\subsubsection{Baseline Model}
\label{sec:baseline_model}

To assess the parameter bias due to TRCs, we compare waveforms against a baseline model. The baseline model includes Newtonian\footnote{The orbital dynamics are Newtonian, but the NS parameters $(I_a, \omega_{a0})$ are scaled to match a relativistic NS described by the SLy equation of state \cite{Read:09a}.}, linear equilibrium and dynamical tides and the same PP orbit given by $\Psi_\mathrm{pp}$. It does not include the TRCs due to nonlinear hydrodynamics, background spin, or GR. The baseline model can be derived from Eqs. (\ref{eq:Delta_a}-\ref{eq:dyn_tide}) by taking the limits: $\mathcal{A}_\mathrm{nl}\rightarrow 0$, $\Omega_\mathrm{A}\rightarrow 0$, $z\rightarrow 1$, and $\Omega_\mathrm{FD}\rightarrow 0$.

For clarity, we provide the explicit expressions for the baseline model here, with the terms unique to the baseline model denoted by the superscript $\mathrm{(b)}$. The baseline equilibrium mode amplitude is,
\begin{equation} \label{eq:baseline_eq_tide}
    b_{a}^\mathrm{(eq,b)} \simeq \frac{\Delta_{a}^\mathrm{(b)} \omega_{a 0} V_{a}}{(\Delta_{a}^\mathrm{(b)})^{2}-i\left[m_{a} \dot{\omega}-\Delta_{a}^\mathrm{(b)}\left(l_{a}+1\right) \frac{\dot{r}}{r}\right]} \,.
\end{equation}
This expression matches Eq. (\ref{eq:eq_tide}) except for a factor of redshift, $z$, and we've introduced another detuning factor for the baseline model,
\begin{equation} \label{eq:baseline_Delta_a}
    \Delta_{a}^\mathrm{(b)} \coloneq \omega_{a0} -m_{a} \omega \,.
\end{equation}
For the baseline model, resonance occurs when this detuning factor equals zero, $\Delta_a^\mathrm{(b)}(\omega_\mathrm{r}^\mathrm{(b)})=0$, which reduces to  $\omega_\mathrm{r}^\mathrm{(b)}= \omega_{a0}/m_a$. Hence the baseline model resonance occurs at the non-spinning f-mode frequency; there are no shifts to the resonance due to background spin, nonlinear hydrodynamics, or GR.

The dynamical tide mode amplitude for the baseline model becomes,
\begin{equation} \label{eq:baseline_dyn_tide}
    c_{a}^{(\mathrm{dyn,b})} \simeq \frac{\,\omega_{a 0} V_{a,2m,\mathrm{r}}}{\sqrt{\dot{\omega}_{\mathrm{pp},\mathrm{r}}}} F(u) \,,
\end{equation}
which matches Eq. (\ref{eq:dyn_tide}) except for a factor of redshift, $z$. The only other difference is evaluating the frequency-dependent terms at the resonance for the baseline model, $\omega_\mathrm{r}^\mathrm{(b)}$. 

Regarding the background spin case, the baseline model does include spin in the PP contribution to the orbit via the dimensionless spin parameters, $\chi_+$ and $\chi_-$, which are defined at the end of \S\ref{sec:intro}. This ensures the parameter bias against the baseline model is due to only the background spin TRC.

\subsection{Comparison with Low-Frequency Expansion in General Relativity}
\label{sec:low_frequency_expansion}

We now compare our nonlinear-tides model derived using Newtonian gravity to the results of \cite{Pitre:25} from their relativistic treatment in the low-frequency limit. They present a tidal response function in Lorentzian form with a modified resonance frequency, $\omega_*$, in equation (1.8). It can be related to our tidal amplitude via the effective Love number for $l=m=2$, defined in eq. (43) of \cite{Yu:25a}, which combines the tidal amplitude over positive- and negative-frequency modes denoted by the subscripts $+$ and $-$,
\begin{equation} \label{eq:tidal_response}
    \frac{k_{2m}}{k_{20}} \simeq \frac{1}{2} \left( \frac{b_{a+}}{V_a} + \frac{b_{a-}}{V_a} \right) + \frac{1}{4} \sum_{a,b}^{m_a + m_b + m = 0} \left( \frac{J_{ab,2m} b_a^{*} b_b^{*}}{V_a} \right) \,,
\end{equation}
where $b_{a\pm}$ are the tidal mode amplitudes with $\{l_a,m_a\}=\{2,m\}$ to conserve angular momentum. The nonlinear correction term (the last term on the right side) comes from eq. (C9) in \cite{Yu:23a}.

To compare results in the low-frequency regime away from resonance, $\omega^2\ll\omega_{a0}^2$, we use the zeroth order solution to the equilibrium tide amplitude with only the nonlinear tidal effects. Combining Eqs. (\ref{eq:b_a0}, \ref{eq:Delta_a}) returns,
\begin{equation} \label{eq:b_a0_pm}
    b_{a\pm} \simeq \frac{\omega_{a0} V_a +\mathcal{C}_\mathrm{nl}}{\omega_{a0}+\mathcal{A}_\mathrm{nl}\omega_{a0}\,\omega^2/\omega_\mathrm{ref}^2\mp m_a\omega} \,.
\end{equation}
The nonlinear coefficient, $\mathcal{C}_\mathrm{nl}$, is introduced to represent the nonlinear corrections in the numerator (corrections to tidal driving; see eq. 17 from \cite{Yu:23a}) that have been dropped in this paper to focus on the denominator corrections affecting the resonance (TRCs).

Using $\omega^2\ll \omega_{a0}^2$ with Eq. (\ref{eq:b_a0_pm}) combined with Eq. (\ref{eq:tidal_response}), we recover a Lorentzian form for the effective Love number which can be expanded into powers of $\omega/\omega_{a0}$,
\begin{eqnarray} 
    \frac{k_{2m}}{k_{20}} &=& \frac{1+ \mathcal{B}_\mathrm{nl}\frac{\omega^2}{\omega_\mathrm{ref}^2} }{1+ 2\mathcal{A}_\mathrm{nl}\frac{\omega^2}{\omega_\mathrm{ref}^2} -m_a^2\frac{\omega^2}{\omega_{a0}^2}} \nonumber \\ 
    &\simeq& 1 + m_a^2 \frac{\omega^2}{\omega_{a0}^2} + (\mathcal{B}_\mathrm{nl}-2\mathcal{A}_\mathrm{nl})\frac{\omega_{a0}^2}{\omega_\mathrm{ref}^2}\frac{\omega^2}{\omega_{a0}^2}\,.
    \label{eq:kappa_2m}
\end{eqnarray}
The numerator term, $\mathcal{B}_\mathrm{nl}$, represents the leading order nonlinear tidal corrections to the effective Love number magnitude. It has absorbed $\mathcal{C}_\mathrm{nl}$ from Eq. (\ref{eq:b_a0_pm}), the $J_{ab,2m}$ terms from Eq. (\ref{eq:tidal_response}), and a factor of $\mathcal{A}_\mathrm{nl}$ from combining the positive and negative frequency terms. The $\mathcal{A}_\mathrm{nl}$ term in the denominator represents the leading order nonlinear tidal correction to the resonance frequency. Note the linear FF correction (the $m_a^2$ term) and nonlinear tidal terms have the same frequency dependence in the denominator of the Lorentzian, $\sim\omega^2$. As a result, the nonlinear TRC is formally as important as the the linear FF corrections studied in \cite{Hinderer:16, Steinhoff:16}, as suggested by \cite{Yu:23a}. Furthermore, in the low-frequency expansion, the two effects are degenerate, leading to biases when the $m_a^2$ term is assumed to be the only correction at $\omega^2$ (see later in Fig. \ref{fig:corner-Anl}).

The second line expansion in Eq. (\ref{eq:kappa_2m}) allows a direct comparison with eq. (8.8) of \cite{Pitre:25}. The first term in the expansion maps to their $k_2$, the static tidal constant; the $m_a^2/\omega_{a0}^2 $ term maps to their $4(\ddot{k}_2/k_2)(M_{\rm A}/R_{\rm A}^3)$ term, accounting for the FF correction of the tidal interaction; and the $(\mathcal{B}_\mathrm{nl}-2\mathcal{A}_\mathrm{nl}) (\omega_{a0}^2/\omega_{\rm ref}^2)$ term maps to their $(p_2/k_2) (M_{\rm B}/M_{\rm tot})$ term, the nonlinear correction.\footnote{
Specifically, $p_2 = -(8\pi/5)\sqrt{\pi/5} I_a^2 (3J_2 + 4\kappa_2I_a)$, $\mathcal{A}_{\rm nl}=\sqrt{\pi/5} (J_2 + 4\kappa_2I_a)(M_{\rm B}/M_{\rm tot})(\omega_{\rm ref}/\omega_{\rm a0})^2$, and $\mathcal{B}_{\rm nl} = -2\sqrt{\pi/5} J_2 (M_{\rm B}/M_{\rm tot})(\omega_{\rm ref}/\omega_{\rm a0})^2$.
These mappings illustrate how the nonlinear corrections contain two degrees of freedom represented by $J_{ab,lm}$ and $\kappa_{abc}$, and the nonlinear coefficients, $\mathcal{A}_\mathrm{nl}$, $\mathcal{B}_\mathrm{nl}$, and $\mathcal{C}_\mathrm{nl}$ are functions of them.}

This expansion also shows the nonlinear TRC entering at the same order (formally 8th post-Newtonian order) as the FF correction to the linear tide (dynamical tide in the convention of \cite{Hinderer:16}), emphasizing the importance of including the nonlinear tides for both the Newtonian and relativistic treatments. It is common practice when modeling tides in GW waveforms to assume only linear tides are significant and use the universal relation to replace $\omega_{a0}$ with tidal deformability in the $m_a^2$ term in Eq. (\ref{eq:kappa_2m}). However the nonlinear term effectively shifts the resonance frequency on the order of ten percent, leading to biases in the inference when the linear universal relation is assumed;
see Figure \ref{fig:corner-Anl}.

Our analysis suggests that the choice of \cite{Pitre:25} to extend the domain of validity of their expansion into a Lorentzian form (see their eq. 8.9), 
\begin{equation}
    \left(\frac{k_{2m}}{k_{20}}\right)^{\rm (PP25)} = \frac{1}{1+ (2\mathcal{A}_\mathrm{nl} - \mathcal{B}_\mathrm{nl}) \frac{\omega^2}{\omega_\mathrm{ref}^2} -m_a^2\frac{\omega^2}{\omega_{a0}^2}}
\end{equation}
is inaccurate, as the nonlinear hydrodynamics corrects both the denominator (the $\mathcal{A}_{\rm nl}$ term) and numerator (the $\mathcal{B}_{\rm nl}$ term) of Eq. (\ref{eq:kappa_2m}). When expanded, both terms contribute to their $p_2$. This further suggests that the effective f-mode frequency $\omega_\ast$ defined in \cite{Pitre:25} (their eq. 1.10) is inaccurate, as it incorrectly absorbs the $\mathcal{B}_{\rm nl}$ term. We instead argue that the effective frequency should be defined as 
\begin{equation}
    \frac{\omega_\ast}{\omega_{a0}} = \frac{1}{1-\frac{\mathcal{A}_{\rm nl}}{4}\frac{\omega_{a0}^2}{\omega_{\rm ref}^2} },
    \label{eq:omega_ast}
\end{equation}
which leads to an effective frequency $\sim 10\%$ lower than its linear value, slightly smaller than the $14\%$ reported by \cite{Pitre:25} for the Newtonian case ($M_{\rm A}/R_{\rm A}=0$). Figure \ref{fig:eff_love} shows the \cite{Pitre:25} resummation result for a Newtonian, $\Gamma=2$ polytrope (green-dotted line), which overestimates the tidal resonance compared to the numerical solution.

\subsection{Effective Love Number}
\label{sec:Eff_Love}

Next we compare our formulation of tidal resonance to the effective Love number approach. A common method for including tidal effects beyond the adiabatic limit is to replace the adiabatic Love number with an effective one, $\Lambda\rightarrow\Lambda_\mathrm{eff}(\omega)$ in the interaction potential; see the description above eq. 6 of \cite{Hinderer:16}. The effective Love number absorbs the orbital frequency dependence due to tidal resonance, thus extending the domain of validity of $Q_{ij}=\Lambda_\mathrm{eff}\mathcal{E}_{ij}$ in the low-frequency regime. Our formulation illustrates some of the limitations of the effective Love number method at higher frequencies close to merger.

First is the lack of tidal torque, a consequence of a purely real effective Love number \cite{Yu:25a}.  The missing torque component can be traced back to replacing the mass quadrupole with the effective Love number in the interaction Hamiltonian. This replacement results in the contraction of the tidal potential with itself, thus eliminating $\phi$ from the interaction Hamiltonian,
\begin{align} \label{eq:H_int}
    H_\mathrm{int}(r, \phi, Q_{ij})& = - \frac{1}{2} Q_{ij} \mathcal{E}^{ij} (r,\phi), \nonumber \\
    \rightarrow H_{\rm int, wrong}(r)&=- \frac{1}{2}\Lambda_\mathrm{eff}(r) \mathcal{E}_{ij}(r,\phi) \mathcal{E}^{ij}(r,\phi) \propto \Lambda_\mathrm{eff}(r) \mathcal{E}^2(r) \,,
\end{align}
where $\mathcal{E}(r)=M_{\rm B}/r^3$. The tidal torque contribution from $\partial H_{\rm int}/\partial \phi$ is lost when there is no $\phi$-dependence in the interaction Hamiltonian used to derive equations of motion. Although the replacement preserves the radial dependence which dominates at low frequencies, the tidal torque becomes significant at higher frequencies and overtakes the radial contribution, see figure 3 in \cite{Yu:24a}, and even without spin \cite{Yu:25a}. 

An alternative way to see the limitation of a single effective Love number is to compare the frequency dependence of the energies. We use the tidal correction to the equilibrium energy ($\Delta E$) to calculate the tidal dephasing which is comprised of the tidal interaction energy ($E_\mathrm{int}\sim b_a$, see Eq. (\ref{eq:Eint_lin})), the tidal mode energy ($E_\mathrm{mode}\sim b_a^2$, see Eq. (\ref{eq:Emode_lin})), and the correction to the orbital energy ($\Delta E_\mathrm{orb}\sim b_a$, see Eq. (\ref{eq:Delta-r})). Both the orbital energy correction and tidal interaction energy go as the tidal mode amplitude -- a Lorentzian dependence before resonance. However, the tidal mode energy goes as the tidal mode amplitude squared. Consequently, the equilibrium energy correction has both Lorentzian and Lorentzian-squared dependence on frequency. This is evident at the linear order of $\Delta E$ in eq. 50 of \cite{Yu:23a} (denoted with subscript ``$\mathrm{eq}$''). The two distinct dependencies on frequency cannot be wholly represented by a single effective Love number introduced to preserve the radial interaction, thus requiring an additional effective Love number that accounts for the Lorentzian-squared component.

Furthermore, the tidal corrections to the GW radiation require yet another effective Love number to capture their frequency dependence, see eq. (147) in \cite{Yu:25a}. This is illustrated by comparing the two GW radiation corrections. The correction due to the modified orbit ($\Delta\dot{E}_{r-\omega}\sim\Delta r$, see Eq. (\ref{eq:Delta-r})) has contributions to all three $l=2$ modes, $m=\{2,0,-2\}$. However, the correction due to NS-orbit coupling (see Eq. (\ref{eq:Edot_ns-orb})) is proportional to $m$, thus only contributing to two modes, $m=\{2,-2\}$. \cite{Yu:25a} show how to define an effective Love number for GW radiation ($\kappa_{\mathrm{eff},h}$) in terms of the effective Love number that preserves the radial interaction.

\subsection{Canonical Neutron Star}
\label{sec:canonical_NS}

Assuming all NSs follow the same EoS, we account for NS mass variability as a perturbation away from a canonical NS. Following \cite{Pratten:22}, we choose our canonical NS to have a mass of $M_\mathrm{cano}=1.33 M_{\odot}$, and a radius of $R_\mathrm{cano}=11.7\times10^5$ cm. We assume that the radii of all NSs are the same at $R=R_\mathrm{cano}$, which approximates NSs as $\Gamma=2$ Newtonian polytropes. We similarly use the polytrope approximation for the scaling of other properties, specifically tidal deformability, f-mode frequency, and nonlinear couplings. This means the Love number, $k_2$, the ratio between f-mode frequency and the NS's dynamical frequency, $\omega_{a0}/\sqrt{M_\mathrm{A}/R^3}$, as well as the nonlinear coupling coefficients $J_2$ and $\kappa_2$ entering our $\mathcal{A}_{\rm nl}$, are all approximated as constants. Nonetheless, we shift the canonical values of $k_2$ and $\omega_{a0}$ so that they are consistent with a relativistic NS with the SLy EoS. While crude, our treatment should be sufficient for this proof-of-principle study.
The canonical values can then be converted to the values for individual NSs using their mass dependence.

Using the SLy EoS, we deduce a dimensionless tidal deformability of $\bar{\lambda}_\mathrm{cano}=400$. This can be related to the tidal love number as $k_\mathrm{cano}=1.5 \bar{\lambda}_\mathrm{cano}(M_\mathrm{cano}/R)^5=0.08$. To calculate the tidal deformability of a NS with mass $M_\mathrm{A}$, the canonical values can be used,
\begin{equation}
    \bar{\lambda}_\mathrm{A} = \left(\frac{M_\mathrm{cano}}{M_\mathrm{A}}\right)^5 \bar{\lambda}_\mathrm{cano} \,,
\end{equation}
where both NS radii are assumed to be equal.

To find the f-mode frequency of a canonical NS, we use the tidal deformability and the universal relation covered in the next section \S\ref{sec:UR}, giving $\bar{\omega}_\mathrm{a0,cano}=0.0786$.
To calculate the f-mode frequency of a NS with mass $M_\mathrm{A}$, the canonical values can be used,
\begin{equation}
    \bar{\omega}_\mathrm{a0,A} = \left(\frac{M_\mathrm{A}}{M_\mathrm{cano}}\right)^{3/2} \bar{\omega}_\mathrm{a0,cano} \,.
\end{equation}

As mentioned in \S\ref{sec:nonlinear_tides} we use a nonlinear TRC magnitude of $\mathcal{A}_\mathrm{nl,cano}=-0.03$ at a reference GW frequency of 1000 Hz. For a NS with mass $M_\mathrm{A}$, the mass-adjusted nonlinear TRC magnitude is
\begin{equation}
    \mathcal{A}_\mathrm{nl,A} = 2\left(\frac{M_\mathrm{cano}}{M_\mathrm{A}}\right)\left(1- \frac{M_\mathrm{A}}{M_{\mathrm{tot}}}\right) \mathcal{A}_\mathrm{nl,cano}\,.
\end{equation}

\subsection{Universal Relations}
\label{sec:UR}

Certain relations between global NS parameters have been shown to be approximately independent of the NS EoS. Thus such universal relations can be used to reduce the number of free parameters when modeling data without knowing the internal physics that determine the EoS. These relations are often referred to as the I-Love-Q relations and have been extended to other physical parameters, such as the f-mode frequency. 

For our work we use two approximately universal relations: the I-Love relation between the stellar moment of inertia and quadrupolar tidal deformability (spin-induced) from \cite{Yagi:17}, and the Love-$\omega_{a0}$ relations between the f-mode frequency and quadrupolar tidal deformability from \cite{Chan:14} (see also, \cite{Saes:25}). These relations hold for various realistic EoSs and polytropes, with a fractional error below 0.01. 

We use the Love-$\omega_{a0}$ relations to calculate the f-mode frequency, $\bar{\omega}_{a0}$, for a given tidal deformability, $\bar{\lambda}$, thus reducing the number of free parameters in the model. 
For the background spin case, we use the I-Love relations to calculate the dimensionless moment of inertia, $\bar{I}$, from the tidal deformability, $\bar{\lambda}$, which can be combined with the dimensionless spin, $\chi_\mathrm{A}$ to get the spin frequency, $\Omega_\mathrm{A}=\chi_\mathrm{A}/\bar{I}_\mathrm{A}$.

Both relations can be represented with the same functional form:
\begin{equation} \label{eq:UR}
    \bar{y}=\bar{a}_{0}+\bar{a}_{1} \ln\bar{\lambda}+\bar{a}_{2} (\ln\bar{\lambda})^{2}+\bar{a}_{3} (\ln\bar{\lambda})^{3}+\bar{a}_{4} (\ln\bar{\lambda})^{4} \,,
\end{equation}
where the $\bar{a}_i$ are best-fit coefficients for realistic NSs and $\bar{y}$ represents the parameter related to $\bar{\lambda}$; both are given in Table \ref{tab:UR_coefficients}. 

\begin{table}
    \centering
    \begin{tabular}{cccccc}           \toprule
        $\bar{y}$ & $\bar{a}_{0}$ & $\bar{a}_{1}$ & $\bar{a}_{2}$ & $\bar{a}_{3}$ & $\bar{a}_{4}$ \\        \midrule
        $\ln\bar{I}$ & 1.496 & 0.05951 & 0.02238 & $-6.953\times 10^{-4}$ & $8.345\times 10^{-6}$ \\
        $\bar{\omega}_{a0}$& $0.1820$ & $-6.836\times 10^{-3}$ & $-4.196\times 10^{-3}$ & $5.215\times 10^{-4}$ & $-1.857\times 10^{-5}$ \\ \bottomrule
    \end{tabular}
    \caption{The best fit coefficients for the universal relations between $\bar{I}-\bar{\lambda}$ (I-Love) and $\bar{\omega}_{a0}-\bar{\lambda}$ (Love-$\omega_{a0}$) given in Eq. (\ref{eq:UR}). These values come from table 1 of \cite{Yagi:17} and table 1 of \cite{Chan:14}.}
    \label{tab:UR_coefficients}
\end{table}

\subsection{Waveform Mismatch}
\label{sec:waveform_mismatch}

To assess the detectability of these TRCs, we use waveform mismatch to quantify the SNR required to discern between waveforms with and without the TRCs. We calculate the mismatch between two waveforms $h_1$ and $h_2$ as
\begin{equation}
    \epsilon \coloneq 1 - \frac{\langle h_1| h_2\rangle}{\sqrt{\langle h_1| h_1\rangle \langle h_2| h_2\rangle}} \,,
\end{equation}
where the angle brackets denote an inner product,
\begin{equation}
    \langle h_1|h_2\rangle \coloneq 2 \int_0^\infty \frac{\tilde{h}_1^* \tilde{h}_2 + \tilde{h}_1 \tilde{h}_2^*}{S_n(f_\mathrm{gw})} df_\mathrm{gw} \,.
\end{equation}
The detector sensitivity curve is denoted by $S_n$.

The lowest threshold SNR needed to discern between two waveforms can be estimated from waveform mismatch as
\begin{equation}
    \rho_\mathrm{th} \simeq (2\epsilon)^{-1/2} \,.
    \label{eq:rho_th}
\end{equation}

We marginalize over the time and phase of coalescence by subtracting out the linear frequency dependence of the waveforms. Results are presented in Table \ref{tab:wm_results}.

\subsection{Parameter Estimation}
\label{sec:parameter_estimation}

For the HMC simulations, we optimize the log-likelihood,
\begin{equation} \label{eq:loglike}
    \ell \coloneq -\frac{1}{2} \int_0^\infty \frac{\left|\tilde{h}_\mathrm{rec}- \tilde{h}_\mathrm{inj}\right|^2}{S_n(f_\mathrm{gw})} df_\mathrm{gw} \,,
\end{equation}
which quantifies the distance between two frequency-domain waveforms -- the recovered waveform, $\tilde{h}_\mathrm{rec}$, which has free parameters optimized by the HMC algorithm; and the injected waveform, $\tilde{h}_\mathrm{inj}$, which is a simulated observed merger -- for a given detector sensitivity curve, $S_n$.

We integrate over frequency up to the approximate GW frequency at coalescence, $f_
\mathrm{c}$, which we estimate as,
\begin{equation}
    (\pi f_\mathrm{c})^2 \simeq \frac{M_\mathrm{tot}}{(1.05(R_\mathrm{A}+R_\mathrm{B}))^3} = \frac{M_\mathrm{tot}}{(2.1R)^3}
    \label{eq:fc}
\end{equation}
where the factor 1.05 is chosen to give a frequency close to coalescence. For the baseline model parameters, $f_\mathrm{c}=1553$ Hz. Note the mass dependence leads to a different coalescence frequency for the injected and recovery waveforms. Since we terminate a waveform at coalescence by setting it to zero for $f>f_\mathrm{c}$, we select the lower $f_\mathrm{c}$ between the injected and recovered waveforms.

We drop the magnitudes of the waveforms in the log-likelihood, keeping only the cross term. Thus Eq. (\ref{eq:loglike}) reduces to, 
\begin{equation} \label{eq:loglike-simplified}
    \ell = \frac{\rho^2}{\bar{\rho}^2} \int_0^{f_\mathrm{c}} \frac{\mathrm{Re}(\tilde{h}_\mathrm{rec}^* \tilde{h}_\mathrm{inj})}{S_n(f_\mathrm{gw})} df_\mathrm{gw} \,,
\end{equation}
where $\rho$ is the SNR of the event, and $\bar{\rho}$ is calculated to normalize the waveform SNR using eq. 24 in \cite{Cutler:94},
\begin{equation}
    \bar{\rho}^2 = 4 \int_0^{f_\mathrm{c}} \frac{|\tilde{h}_\mathrm{inj}|^2}{S_n(f_\mathrm{gw})} df_\mathrm{gw} \,.
\end{equation}

\section{Results}
\label{sec:results}

\subsection{Detectability}
\label{sec:detectability}

\begin{table}
    \centering
    \begin{tabular}{lccc}           \toprule
        TRC& CE & ET & A\# \\        \midrule
        $\mathcal{A}_\mathrm{nl}=-(0.03\pm0.01)$& $85^{-37}_{+65}$ & $118^{-51}_{+90}$ & $62^{-27}_{+49}$ \\
        $\Omega_\mathrm{A}/2\pi=-(200\pm100)$ Hz& $61^{-27}_{+82}$ & $83^{-36}_{+111}$ & $46^{-20}_{+62}$ \\
        $\Omega_\mathrm{A}/2\pi=200\pm100$ Hz& $107^{-27}_{+82}$ & $147^{-36}_{+111}$ & $82^{-20}_{+62}$ \\ 
        GR & 215 & 300 & 158 \\     \bottomrule
    \end{tabular}
    \caption{SNR thresholds for waveform mismatch between a model with a TRC and a baseline model without; see Eq. (\ref{eq:rho_th}). The TRCs are nonlinear hydrodynamics , $\mathcal{A}_\mathrm{nl}$; anti-aligned (negative) and aligned (positive) NS background spin, $\Omega_\mathrm{A}$; and relativistic redshift and frame dragging, GR. Waveform mismatch is calculated for three detectors' sensitivities, Cosmic Explorer (CE), Einstein Telescope (ET), and Advanced LIGO A\# (A\#), and marginalized over the time and phase of coalescence.}
    \label{tab:wm_results}
\end{table}

To estimate the detectability of the three types of TRCs, we use waveform mismatch to calculate the lowest SNR needed (Eq. \ref{eq:rho_th}) to discern between two waveforms: a baseline waveform and a waveform with the specified TRC. The \textbf{baseline waveform} is calculated for a BNS system using Newtonian, linear tides, see \S\ref{sec:baseline_model}. For the baseline waveform, we take the NS masses to be $M_\mathrm{A}=1.29 M_\odot$ and $M_\mathrm{B}=1.37 M_\odot$, resulting in a chirp mass and mass ratio of $\mathcal{M}=1.157 M_\odot$ and $q=0.942$. We also use the canonical parameter values from \S\ref{sec:canonical_NS} for radius $R=11.7\times 10^5$ cm, tidal deformability $\bar{\lambda}=400$, and dimensionless f-mode frequency $\bar{\omega}_{a0}=0.0786$. Time and phase of coalescence are marginalized over by subtracting out a linear fit of the difference between the two waveform phases. We do not include any TRCs in the baseline model, hence $\mathcal{A}_{\rm nl}=\Omega_{\rm A}=\Omega_{\rm B}=0$. For spinning systems, spin enters the waveform in two ways: the PP sector via $\chi_+$ and $\chi_-$ and the tidal sector via $\Omega_{\rm A}$ and $\Omega_{\rm B}$. To isolate the impact of the background spin TRC, we include the PP contribution from spin in the baseline model by matching the baseline's $\chi_+$ and $\chi_-$ values to the TRC waveform's $\chi_+$ and $\chi_-$ values which are calculated from the spin TRC values, $\Omega_{\rm A}$ and $\Omega_{\rm B}$. The \textbf{TRC waveform} has the same parameters as the baseline waveform except for one of the three TRCs -- the nonlinear hydrodynamics TRC, the background spin TRC due to Doppler and Coriolis effects, or the relativistic TRC from redshift and frame dragging.

The nonlinear hydrodynamics and background spin (aligned and anti-aligned) TRCs are evaluated at three values each, denoted by the $\pm$. The difference in phase between the TRC waveforms and baseline waveforms is presented in Figure \ref{fig:f-mode effects}. We evaluate the SNR threshold by integrating over frequency up to merger which we estimate at a separation of $r_\mathrm{c}=2.1 R$, see Eq. (\ref{eq:fc}). This gives a coalescence frequency of $f_\mathrm{c}=1553$ Hz for the baseline parameters. The waveforms are evaluated at frequencies from 100 Hz to $f_{\rm c}$ with a linear spacing of 1 Hz. We verify that the down-sampling of frequency is sufficient to capture the phase evolution in ranges where the tide is significant. We present detectability results in Table \ref{tab:wm_results} for three detector sensitivities: Cosmic Explorer\footnote{https://dcc.cosmicexplorer.org/CE-T2000017/public} (CE) \cite{Evans:17}, Einstein Telescope\footnote{https://www.et-gw.eu/index.php/etsensitivities} (ET) \cite{Punturo:10}, and Advanced LIGO A\#\footnote{https://dcc.ligo.org/LIGO-T2300041/public} (A\#).

For the nonlinear TRC, we find SNR thresholds of the 30s to over 200 are needed to detect them. This range accounts for the variability of $\mathcal{A}_\mathrm{nl}$ due to EoS dependence and GR corrections. The frequency dependence of the nonlinear TRC results in a waveform that stays closer to the baseline for most of the inspiral, while the large deviation away from the baseline occurs close to merger.

For background spin, an anti-aligned spin shifts the resonance to lower frequency, thus amplifying the resonance in the waveform. As a result, anti-aligned spin has the lower SNR thresholds than aligned spin. Both aligned and anti-aligned spins result in an SNR threshold on the orders of  $\mathcal{O}$(10-100), assuming a spin frequency in the hundreds of Hertz. Because spin shifts the resonance by a constant amount, the resulting waveform deviates from the baseline waveform earlier in the inspiral than the other two TRCs, as shown in Figure \ref{fig:f-mode effects}.

For detecting the GR TRC, an SNR in the hundreds is required. As mentioned in \cite{Steinhoff:21}, the redshift and frame-dragging effects counteract each other, with frame dragging dominating and pushing the resonance to higher frequency. These effects are also frequency dependent, further pushing their significant contributions to higher frequency and closer to merger.

We find Advanced LIGO (A\#) gives the lowest SNR thresholds of the three detectors due to its higher sensitivity at higher frequencies -- where the TRCs manifest -- relative to its entire sensitivity curve; Waveform mismatch is insensitive to a detector's overall sensitivity. The CE detector, with the highest overall sensitivity, will detect events with higher SNR.

\subsection{Parameter Bias}
\label{sec:parameter_bias}

To assess the impact TRCs have on systematic bias in parameter estimation, we use a baseline model (see \S\ref{sec:baseline_model}) without TRCs to recover posterior probability distributions by fitting an injected waveform that includes one of the three TRCs. We then quantify the bias of the recovered estimate for tidal deformability -- the primary parameter used to infer the NS EoS. For comparison, we also use a recovery model which includes the injected TRC to recover the true parameter values from the injected waveform. 

The baseline model uses the baseline parameter values presented in \S\ref{sec:detectability}, with the free parameters $\{q, t_\mathrm{c}, \varphi_\mathrm{c}, \bar{\lambda}\}$. Note that $\omega_{a0}$ is reduced via its linear universal relation with $\bar{\lambda}$. For the background spin TRC, the dimensionless spin parameters, $\chi_\pm$, are also included as free parameters in the baseline model. The injected waveform uses the same parameter values as the baseline model, but one of the three TRCs is included. The complete model includes the three TRCs in three different ways: for nonlinear hydrodynamics it includes $\mathcal{A}_\mathrm{nl}$ as a free parameter; for background spin it constrains $\Omega_\mathrm{A}$ using the dimensionless spin parameters, $\chi_\pm$, and the NS moment of inertia calculated via the I-Love relation; and for GR it includes the redshift and frame-dragging corrections.

We use HMC simulations to estimate parameter values for an injected, noise-free\footnote{Note that Gaussian-noise realizations may shift the resulting posteriors by $\sim\sigma$ but average out to the noise-free realization.} waveform, fitted against either the baseline model or TRC model using the log-likelihood defined in Eq. (\ref{eq:loglike-simplified}) and the CE sensitivity curve. The values of the free parameters in the baseline and TRC models are sampled from uniform priors with bounds given in Table \ref{tab:prior_bounds}. The HMC simulations are initialized at the injection values and iterate over a few thousand warm-up samples followed by the 30,000 samples used to generate posteriors. We use the No U-Turn Sampler (NUTS) version of HMC as implemented in the NumPyro python package \cite{phanComposable:19,BinghamPyro:19}. The python library JAX \cite{JAXGithub:18} accelerates the HMC computations.

\begin{table}
    \centering
    \begin{tabular}{lccc}           \toprule
        Parameter & Lower bound & Upper bound & Units \\        \midrule
        $q$ & 0.7 & 0.999 & - \\
        $t_\mathrm{c}$ & -1 & 1 & milliseconds \\
        $\varphi_\mathrm{c}$ & $-\pi$ & $\pi$ & radians \\
        $\bar{\lambda}$ & 100 & 1300 & - \\
        $\bar{\omega}_{a0}$& 0.04 & 0.12 & - \\
        $\mathcal{A}_\mathrm{nl}$& -0.06 & 0 & - \\
        $\chi_+$, $\chi_-$ & -1 & 1 & - \\     \bottomrule
    \end{tabular}
    \caption{
    Upper and lower bounds of uniform priors for each model parameter used in the HMC simulations. The parameters are mass ratio, time and phase of coalescence, dimensionless tidal deformability, dimensionless non-spinning f-mode frequency, nonlinear TRC parameter, and dimensionless spin parameter combinations.
    The nonlinear tides parameter is restricted to only negative values, except for the HMC simulation plotted in purple in Figure \ref{fig:hist-Anl}, denoted by $\mathcal{A}_\mathrm{nl}^{(\pm)}$.
    }
    \label{tab:prior_bounds}
\end{table}

\subsubsection{Nonlinear Tides}

\begin{figure}
    \centering
    \includegraphics[width=0.5\linewidth]{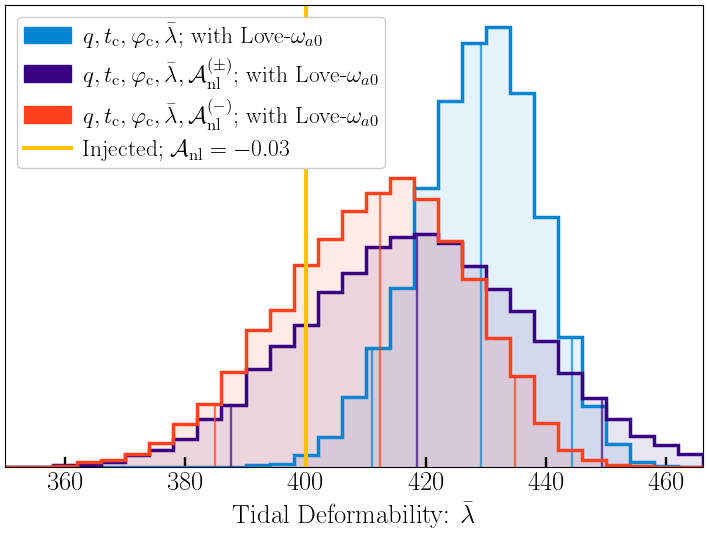}
    \caption{
    Comparison of the recovered posteriors for an injected waveform with the nonlinear TRC, $\mathcal{A}_\mathrm{nl}=-0.03$ and a SNR of 500. 
    The blue posterior is recovered from the baseline model that has the free parameters listed in the legend, which does not include the nonlinear TRC, \textit{i.e.}, with $\mathcal{A}_\mathrm{nl}=0$.
    The purple posterior is recovered from a model that includes the nonlinear TRC with a uniform prior on $\mathcal{A}_{\rm nl} \in [-0.06, 0.06]$.
    The red posterior is recovered from a model that includes the nonlinear TRC whose prior is only negative.
    All three recovery models impose the Love-$\omega_{a0}$ relation.
    The vertical gold line denotes the injected value, $\bar{\lambda}=400$.
    The 5, 50, and 95 percentiles are denoted by vertical lines within each posterior.
    }
    \label{fig:hist-Anl}
\end{figure}

The one-dimensional, marginalized posterior distributions for tidal deformability from fitting an injected waveform (includes the nonlinear TRC, $\mathcal{A}_\mathrm{nl}=-0.03$ and SNR of 500) with the baseline model (blue) and the TRC model (red and purple) are plotted in Figure \ref{fig:hist-Anl}. The purple posterior is from fitting the TRC model with an expanded prior distribution that includes positive $\mathcal{A}_\mathrm{nl}$ values while the red one is restricted to a prior with $\mathcal{A}_{\rm nl}<0$. All three estimations of tidal deformability assume the Love-$\omega_{a0}$ relation, where the non-spinning f-mode frequency is determined by tidal deformability. The baseline model shows a +8\% bias of tidal deformability, with a best-fit value of 432 compared to the injected value of 400. When NS mass varies (see the following subsection on event stacking), the tidal deformability bias varies between +5\% and +9\%. For comparison, \cite{Pratten:22} find roughly +30\% systematic bias of tidal deformability which results from fitting linear tides with a model that only includes adiabatic tides. Such a bias on tidal deformability ultimately leads to $\mathcal{O}$(1km) error when inferring the NS EoS \cite{Pratten:22}. Not including nonlinear tides leads to an additional +8\% bias in the same direction, a bias toward higher tidal deformability.

\begin{figure}
    \centering
    \includegraphics[width=0.5\linewidth]{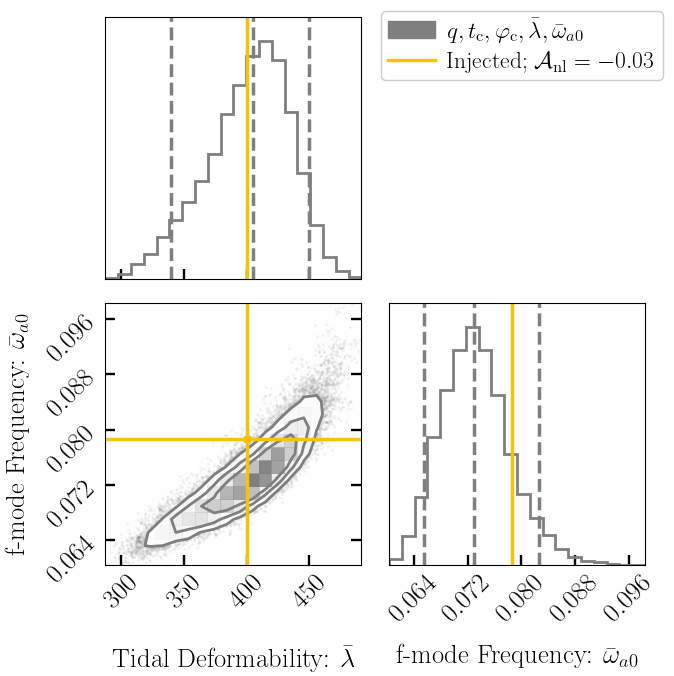}
    \caption{
    Corner plot of the recovered posterior for an injected waveform with the nonlinear TRC, $\mathcal{A}_\mathrm{nl}=-0.03$ and SNR of 500. The gray posterior is recovered with the baseline model which has the free parameters listed in the legend and does not include the nonlinear TRC, $\mathcal{A}_\mathrm{nl}=0$. Unlike Figure \ref{fig:hist-Anl}, the Love-$\omega_{a0}$ relation is not used, allowing the f-mode frequency, $\bar{\omega}_{a0}$, to vary.
    The gold lines are the injected values, $\bar{\lambda}=400$ and $\bar{\omega}_{a0}=0.0786$. The dashed lines denote the 5, 50, and 95 percentiles. The contours in the 2D plot enclose 50\%, 75\%, and 90\% of the samples.
    }
    \label{fig:corner-Anl}
\end{figure}

We also fit the nonlinear-TRC waveform with the baseline model but without the Love-$\omega_{a0}$ relation, leaving $\bar{\omega}_{a0}$ as a free parameter. The one- and two-dimensional posterior plots for tidal deformability and f-mode frequency are shown in Figure \ref{fig:corner-Anl}, with the gold lines denoting the injected values. Without the universal relation, tidal deformability becomes consistent with the injected value, shifted by only -2\%, a best-fit value of 392 versus the true 400. In contrast, the f-mode frequency is biased by -10\%, with a recovered value of 0.0711 versus the injected 0.0786. This shows that not including the nonlinear TRC primarily manifests as bias of the f-mode frequency which is then transferred to tidal deformability when imposing the Love-$\omega_{a0}$ relation. 

\begin{figure}
    \centering
    \includegraphics[width=0.5\linewidth]{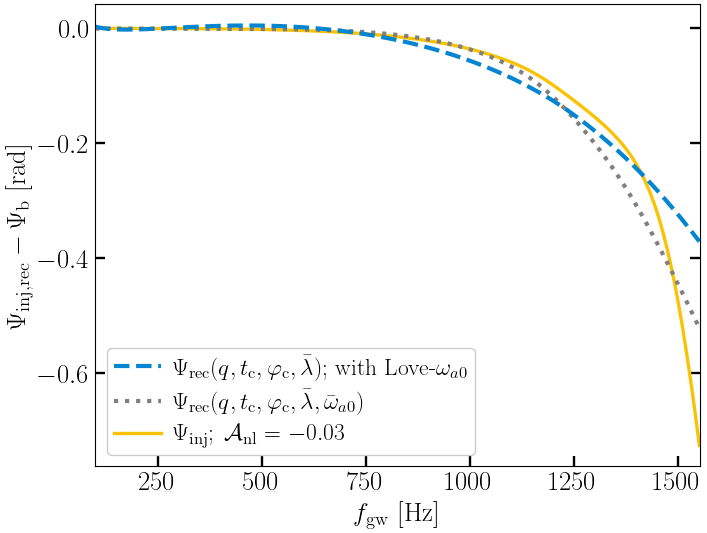}
    \caption{
    The difference between the waveform phase of the baseline model ($\Psi_\mathrm{b}$) and the waveform phases of a.) solid gold: the injected model with the nonlinear TRC ($\mathcal{A}_\mathrm{nl}$); b.) dotted gray: the recovered model with the f-mode frequency ($\bar{\omega}_{a0}$) as a free parameter; c.) dashed blue: the recovered model with the f-mode frequency ($\bar{\omega}_{a0}$) constrained by tidal deformability ($\bar{\lambda}$) and the Love-$\omega_{a0}$ relation.
    Using waveform mismatch, the SNR threshold is 206 between the blue and gold waveforms and 354 between the gray and gold waveforms.
    }
    \label{fig:bw-Anl}
\end{figure}

Figure \ref{fig:bw-Anl} shows the injected (gold) versus recovered waveform phases of the blue and gray posteriors from Figures \ref{fig:hist-Anl} and \ref{fig:corner-Anl}. The baseline waveform, whose phase is denoted by $\Psi_\mathrm{b}$, is evaluated at the injected parameter values, with Newtonian, linear tides but no TRCs. The baseline waveform phase is subtracted out to show deviation away from linear tides. The (dashed) recovered waveform phase from the blue posterior tries to account for the nonlinear TRC with a biased tidal deformability, but it fails to capture the sharp change in phase at high frequency. The (dotted) recovered waveform phase from the gray posterior is able to closely fit the injected waveform to higher frequency, due to the additional degree of freedom, but also fails to capture the sharp change in phase above 1200 Hz. Applying waveform mismatch between the injected and recovered waveforms gives a threshold SNR of 206 between the blue and gold waveforms and a threshold SNR of 354 between the gray and gold waveforms -- an estimate of the minimum SNR needed to discern between the two waveforms. Note the recovered waveform is marginalized over more free parameters than only the time and phase of coalescence as was done for the detectability results. More free parameters typically results in a closer fit to the injected waveform and thus a higher mismatch threshold SNR.

\begin{figure}
\centering
    \begin{subfigure}{0.5\textwidth}
        \centering
        \includegraphics[width=0.95\linewidth]{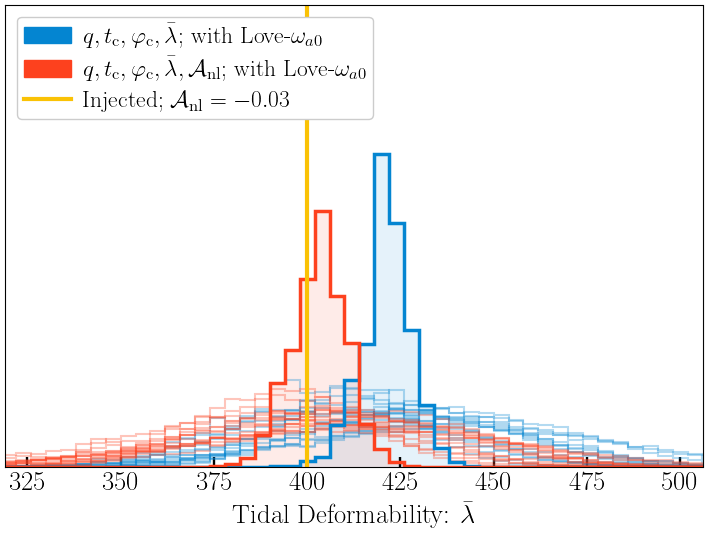}
    \end{subfigure}%
    \begin{subfigure}{0.5\textwidth}
        \centering
        \includegraphics[width=0.95\linewidth]{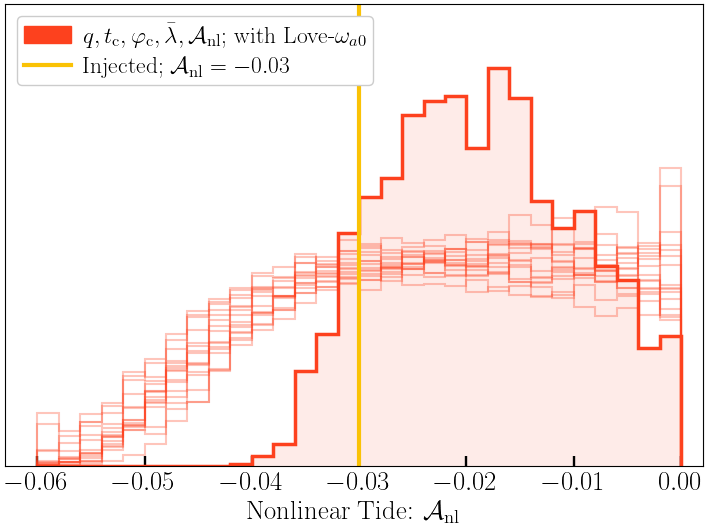}
    \end{subfigure}
    \caption{Comparison of the recovered posteriors, stacked over 16 events, for an injected model with the nonlinear TRC, $\mathcal{A}_\mathrm{nl}=-0.03$. Each event has a SNR of 175 but different NS masses. 
    The blue posteriors are recovered from a model that does not include the nonlinear TRC, $\mathcal{A}_\mathrm{nl}=0$. 
    The red posteriors are recovered from a model that includes the nonlinear TRC as a free parameter, $\mathcal{A}_\mathrm{nl}$. 
    The thin, faint posteriors are for individual events, and the thick, shaded posteriors are the product of stacking the 16 individual event posteriors. 
    The vertical gold lines are the injected values.
    (Left) Stacked posteriors for tidal deformability with an injected value of, $\bar{\lambda}=400$.
    (Right) Stacked posterior for the nonlinear TRC with an injected value of $\mathcal{A}_\mathrm{nl}=-0.03$. }
    \label{fig:histPop-lambda-Anl}
\end{figure}

There is a significant projection effect when marginalizing the five-dimensional red and purple posteriors in Figure \ref{fig:hist-Anl} to get the one-dimensional posteriors for tidal deformability. The mode of the 1D $\bar{\lambda}$ posterior is offset from the mode of the 5D posterior which aligns with the injected value of $\bar{\lambda}=400$. This significant offset is caused by the correlation between $\bar{\lambda}$ and $\mathcal{A}_\mathrm{nl}$ and the skew of $\mathcal{A}_\mathrm{nl}$ toward positive values. Such skew can be seen in the $\mathcal{A}_\mathrm{nl}$ plot of Figure \ref{fig:histPop-lambda-Anl}. The correlation is apparent when comparing the purple and red posteriors in Figure \ref{fig:hist-Anl}; when $\mathcal{A}_\mathrm{nl}$ is restricted to negative values, then the reduced skew in $\mathcal{A}_\mathrm{nl}$ correlates with a shift toward the injected value in the $\bar{\lambda}$ posterior. A result of this offset's dependence on $\mathcal{A}_\mathrm{nl}$, the blue posterior does not see such a projection effect, and the 1D $\bar{\lambda}$ posterior mode aligns more closely with the 4D posterior mode. We also note this projection effect decreases as SNR increases.

\textit{Event Stacking}$\quad$
Although the nonlinear TRC requires large SNR to be detectable in individual events, information from multiple events can be combined to achieve tighter constraints on parameter estimates. To demonstrate this we stack 16 BNS events of the same SNR but each with different NS masses, following \cite{Pratten:22}. For the 16 events, we sample 32 NS masses from the Gaussian distribution of double NS systems with mean 1.33 $M_\odot$ and standard deviation 0.09 $M_\odot$\cite{Ozel:16}. Out of the 32 sampled masses, the lowest is 1.1692 $M_\odot$ and the highest is 1.5265 $M_\odot$. We then sort each pair of NSs such that $M_\mathrm{A}<M_\mathrm{B}$ and calculate the mass ratio and chirp mass, resulting in a mass ratio range of 0.7891 to 0.9965, and a chirp mass range of 1.0299 $M_\odot$ to 1.2901 $M_\odot$. We use the same parameter values as the baseline model for all 16 events except for the NS masses. The injected waveform has the nonlinear TRC with $\mathcal{A}_\mathrm{nl}=-0.03$. We then follow the same parameter estimation process using the HMC method with two recovery models, with and without the nonlinear TRC. Finally we stack the 1D marginalized posteriors by taking the product of counts per bin \cite{Pratten:22}.

The stacked, 1D marginalized posteriors for tidal deformability and nonlinear tides from 16 BNS events with SNR of 175 are shown in Figure \ref{fig:histPop-lambda-Anl}. We find that stacking events reduces the standard deviation in the marginalized posteriors, approximately following the expected one over the square root of the number of events, $\sigma_\mathrm{stacked}\simeq\sigma_\mathrm{event}/\sqrt{N_\mathrm{events}}$ . For 16 stacked events with SNRs of 125, the uncertainty of tidal deformability is constrained enough to discern the inferred bias; the injected value for tidal deformability is at the 12.5 percentile of the stacked baseline posterior. However, the nonlinear TRC parameter is not adequately constrained to negative values with SNRs of 125. For 16 events with SNRs of 175, the injected value is at the 1.25 percentile of the stacked tidal-deformability posterior, and the uncertainty of the nonlinear tides parameter has dropped enough to constrain $\mathcal{A}_\mathrm{nl}$ as nonzero and negative.

We note the projection effect mentioned earlier in this section persists when stacking events. The modes of the 1D, $\bar{\lambda}$ posteriors for individual events are offset from the full posterior mode which persist when stacking posteriors -- unlike the posterior's variance which decreases -- making the offset more pronounced. The projection effect is more pronounced when $\mathcal{A}_\mathrm{nl}$ extends into positive values, such as the purple posterior in Figure \ref{fig:hist-Anl}.

\subsubsection{Neutron Star Background Spin}

\begin{figure}
    \centering
    \begin{subfigure}{0.5\textwidth}
        \centering
        \includegraphics[width=0.95\linewidth]{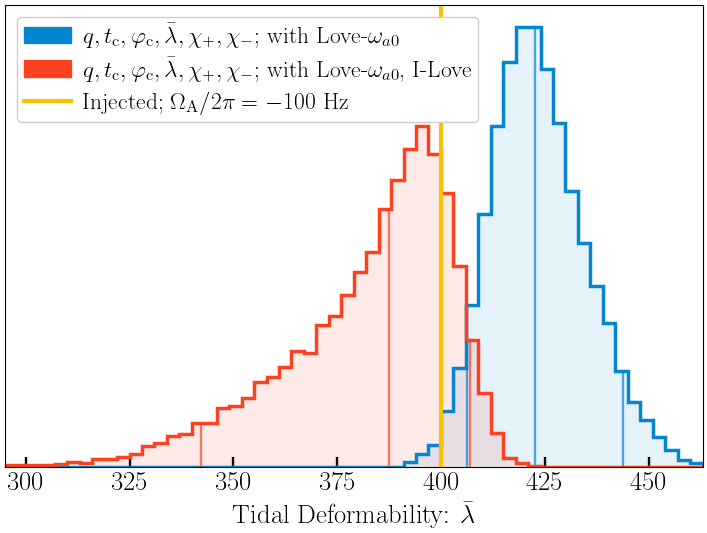}
    \end{subfigure}%
    \begin{subfigure}{0.5\textwidth}
        \centering
        \includegraphics[width=0.95\linewidth]{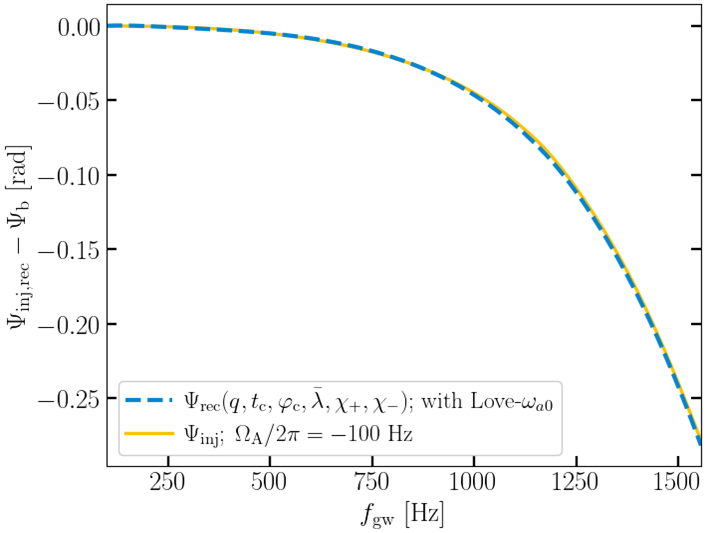}
    \end{subfigure}
    \caption{
    The same as Figures \ref{fig:hist-Anl} and \ref{fig:bw-Anl}, but using an injected waveform with a spinning primary NS, resulting in a background spin TRC of $\Omega_\mathrm{A}/2\pi=(-100$ Hz). The SNR for both posteriors is 1000.
    Both recovery models include spin in the PP contribution to the waveform via $\chi_\pm$, but neither include the background spin TRC as a free parameter. Instead, the red posterior is recovered using a model that constrains the background spin TRC ($\Omega_\mathrm{A}$) using $\chi_\mathrm{A}$ and the I-Love relation, thus enabling the model to recover the injected parameter values.
    The waveform mismatch between the injected and recovered waveforms implies a threshold SNR of 2900 to discern between the two.
    }
    \label{fig:hist-bw-S1}
\end{figure}

The 1D marginalized posteriors for tidal deformability for an injected waveform with the primary NS spinning at -100 Hz (anti-aligned with the orbital angular momentum) and SNR of 1000 are shown in the Figure \ref{fig:hist-bw-S1} left panel. The companion star is not spinning, hence $\Omega_\mathrm{A}=-100$ Hz, $\Omega_\mathrm{B}=0$, and $\chi_+=-0.0250$, $\chi_-=-0.0257$. The baseline (blue) and TRC (red) recovery models both use the same free parameters, listed in the legend, and both assume the Love-$\omega_{a0}$ relation. The dimensionless spin parameters, $\chi_\pm$, are included as free parameters and enter the waveform via $\Psi_\mathrm{pp}$ in Eq. (\ref{eq:waveform-phase}). The baseline recovery model does not include the background spin TRC which is parameterized as $\Omega_\mathrm{A}=0$ \footnote{We treat $\Omega_\mathrm{A}$ and $\chi_\mathrm{A}$ as independent variables. Even though they both represent the same physical characteristic, NS spin, they enter the waveform differently, where $\Omega_\mathrm{A}$ specifies the TRC.}. The TRC recovery model uses the I-Love relation to calculate the background spin TRC parameter via $\Omega_\mathrm{A}=\chi_\mathrm{A} M_\mathrm{A}^2/I_\mathrm{A}$, where $\chi_\mathrm{A}$ is calculated from $\chi_\pm$, and the NS moment of inertia, $I_\mathrm{A}$, is constrained by the tidal deformability and I-Love relation. Thus the TRC recovery model includes the background spin TRC without an additional free parameter.

We find the baseline model recovers a tidal deformability of 426 versus the injected value of 400, a bias of +6\%. This places the recovered value at 2.3 standard deviations away from the true value for an SNR of 1000. The complete model recovers the injected value with a strong skew toward lower tidal deformability. 

The injected waveform and recovered waveform using the baseline (blue) model are compared in the right panel of Figure \ref{fig:hist-bw-S1}. The near perfect match between the injected and recovered waveforms illustrates the degeneracy between the background spin TRC and the f-mode frequency. Evident in Eqs. (\ref{eq:omega-co-rotating}) and (\ref{eq:equil_mode_gov_eq}), the spin parameter $\Omega_\mathrm{A}$ can be absorbed into a redefined effective f-mode frequency, $\omega_{a0}^\mathrm{eff}\coloneq\omega_{a0}+m_a\Omega_\mathrm{A}(1-1/l)$. This degeneracy is transferred to tidal deformability via the Love-$\omega_{a0}$ relation as a bias. The tiny difference between the waveforms is due to the finite sampling of the HMC simulation, where the best-fit sample does not quite hit the exact log-likelihood maximum.

\subsubsection{Redshift and Frame Dragging}

\begin{figure}
    \centering
    \begin{subfigure}{0.5\textwidth}
        \centering
        \includegraphics[width=0.95\linewidth]{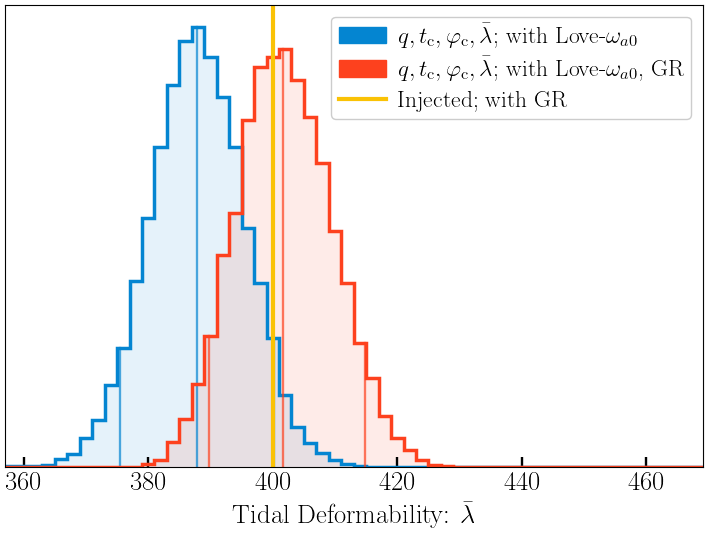}
    \end{subfigure}%
    \begin{subfigure}{0.5\textwidth}
        \centering
        \includegraphics[width=0.95\linewidth]{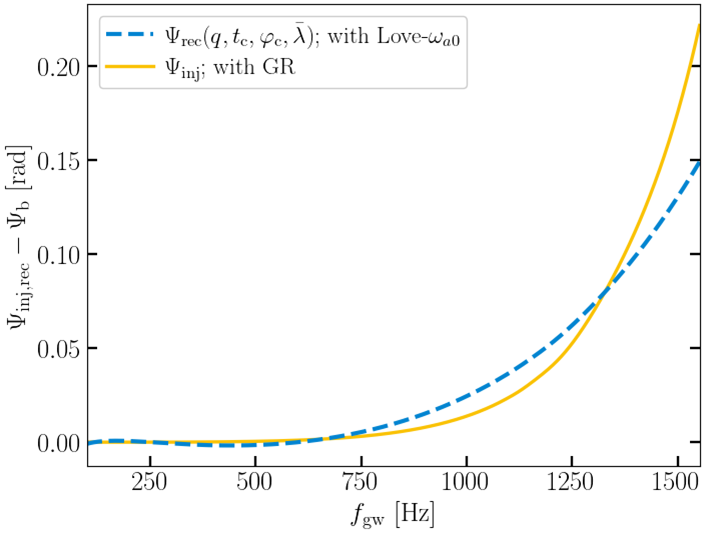}
    \end{subfigure}
    \caption{
    The same as Figures \ref{fig:hist-Anl} and \ref{fig:bw-Anl}, but using an injected waveform with the relativistic TRC. The SNR for both posteriors is 1000. Both recovery models have the same four free parameters and impose the Love-$\omega_{a0}$ relation, but the red posterior includes the relativistic TRC. The recovery model for the blue posterior does not include relativistic effects and cannot fully recover the injected waveform. The best fit waveform from the blue posterior is plotted over the injected waveform in the right panel. 
The mismatch between the injected and recovered waveforms implies a threshold SNR of 610.
    }
    \label{fig:hist-bw-GR}
\end{figure}

The 1D marginalized posteriors for tidal deformability for an injected waveform with redshift and frame-dragging relativistic effects \cite{Steinhoff:21} and SNR of 1000 are shown in the left panel of Figure \ref{fig:hist-bw-GR}. The baseline (blue) and TRC (red) recovery models both use the same free parameters, listed in the legend, and both assume the Love-$\omega_{a0}$ relation. The red TRC model includes relativistic effects and recovers the injected value for tidal deformability, while the blue baseline model does not include GR and recovers a biased estimation for tidal deformability. We find that not including relativistic corrections to the tidal resonance results in a tidal deformability bias of -3\%, recovering a value of 387 versus the injected 400. For an SNR of 1000, the bias is 1.7 standard deviations away from true. 

The recovered waveform phase of the blue posterior is plotted in the right panel of Figure \ref{fig:hist-bw-GR} on top of the injected waveform phase in gold. Since the relativistic corrections push the tidal resonance to higher frequency and result in less resonance amplification of the tidal deformation, then the injected waveform adds phase to the baseline waveform (linear tides). This is opposite to the nonlinear tides and anti-aligned spin cases, which reduced the resonance frequency and consequently the waveform phase. The model with the relativistic TRC primarily biases tidal deformability to lower values to better fit the injected waveform. Due to the lower overall amplitude of relativistic effects on the tidal resonance, the waveform mismatch between the recovered and injected waveforms is larger, at 610.

\section{Conclusions and Discussion}
\label{sec:conclusions}

The key result of our work is the significant systematic bias that is the result of estimating tidal deformability with a model that does not include nonlinear hydrodynamical effects on the f-mode resonance frequency. Our calculations result in an estimated bias of +8\%, indicating the importance of further studying the impact of nonlinear tidal effects with more realistic models. A more detailed analysis should lead to an even more significant bias, as we have chosen to, as a conservative approximation, drop the nonlinear correction to the tidal drive, see Fig. \ref{fig:eff_love}. This tidal deformability bias is in addition to the +30\% bias found by \cite{Pratten:22}, which was shown to bias the inferred NS radius by $\mathcal{O}$(1km). 
According to our detectability results in Table \ref{tab:wm_results} which marginalized over time and phase of coalescence, for the bias to exceed the statistical error from a single detection, an SNR of at least $\sim 100$ is needed (see also Fig. \ref{fig:hist-Anl}). Whereas when additionally marginalizing over mass ratio and tidal deformability, the threshold SNR rises to $\sim 200$ (see Fig. \ref{fig:bw-Anl}). Alternatively, the systematic error exceeds the statistical error when stacking 16 SNR-125 events (Fig. \ref{fig:histPop-lambda-Anl}).
Given a plethora of potential EoSs, it will require a measurement of NS radius with precision of $\mathcal{O}$(0.1km) to constrain the EoS. Such precision is washed out by an $\sim8\%$ systematic bias on tidal deformability. Our work illustrates the importance of including nonlinear hydrodynamical effects for precise measurements of tidal deformability.

We also find the bias on tidal deformability is a direct result of using the Love-$\omega_{a0}$ relation between the tidal deformability and the f-mode frequency. Without the relation, the estimated f-mode frequency takes on the bias from excluding the nonlinear TRC; the biased value being $\sim-10\%$ of the injected value. From the same simulation, tidal deformability is consistent with the injected value. Consequently, using the Love-$\omega_{a0}$ relation transfers the bias from the f-mode frequency to tidal deformability. We recommend abstaining from using the linear Love-$\omega_{a0}$ universal relation for parameter estimation to avoid transferring such f-mode bias to tidal deformability and subsequently the EoS. The linear Love-$\omega_{a0}$ universal relation misses significant nonlinear hydrodynamics as demonstrated by this work and \cite{Yu:23a, Pitre:25, Pani:25}. A nonlinear Love-$\omega_{a0}$ universal relation may be a viable route to include the additional physics while reducing the number of free parameters during inference.

We use waveform mismatch to estimate the threshold SNR needed to detect the TRCs, with results posted in Table \ref{tab:wm_results} for three next-generation detectors: Cosmic Explorer, Einstein Telescope, and Advanced LIGO A\#. We find that a single, loud event needs an SNR of at least $\mathcal{O}$(10-100) for the TRCs to be detectable. The nonlinear TRC starts to become detectable in single events with SNR of at least $\sim$100, depending on the NS EoS and detector. The impact of the background spin TRC on the waveform is detectable at a SNR of $\sim$40 for an anti-aligned spin of -300 Hz, compared to a SNR of $\sim$140 for an anti-aligned spin of -100 Hz. For background spin that is aligned with the orbit, a SNR of $\sim$80 is needed for +300 Hz, versus $\sim$190 for +100 Hz. The GR TRC -- redshift and frame dragging -- requires an SNR of over 200 to be detectable, due in part to redshift counteracting frame dragging \cite{Steinhoff:21}. Keep in mind these detectability estimates are only indicative, and a more detailed analysis is needed to arrive at more accurate thresholds.

Assuming the NS EoS and the nonlinear effects are universal for all NSs, then multiple events can be stacked to improve constraints on parameters. Figure \ref{fig:histPop-lambda-Anl} illustrates how 16 events with different NS masses can be stacked to achieve tighter posteriors for the tidal deformability and nonlinear tides parameters. We find that 16 stacked events with an SNR of 125 each puts the injected, true value outside the 12.5 percentile of the recovery posterior. An SNR of 175 each is needed to stack 16 events and constrain the nonlinear TRC parameter as negative. We find the spread of the stacked posterior approximately goes as one on the square root of the number of stacked events, assuming all events have the same SNR.

We compare our Newtonian treatment of nonlinear tide with the relativistic results of \cite{Pitre:25} by comparing the (radial) effective Love number in the low-frequency expansion. We find a direct mapping between the Newtonian and relativistic nonlinear tidal terms, where both enter at the same order, $(\omega/\omega_{a0})^2$, as the linear FF corrections studied by \cite{Hinderer:16, Steinhoff:16}. The dominant nonlinear hydrodynamical effect is a shift of the effective f-mode frequency, Eq. (\ref{eq:omega_ast}), which lowers the mode frequency by about 10\%. This is smaller compared to the 14\% estimation from \cite{Pitre:25} (see their fig. 4 and our Fig. \ref{fig:eff_love}), as we believe the nonlinear tidal effects should stay separated into corrections to both the numerator and denominator of the Lorentzian description of the effective Love number (Eq. \ref{eq:kappa_2m}; \cite{Yu:23a}). In this analysis, we dropped the numerator effects for simplicity. Even under this conservative approximation (shown in Fig. \ref{fig:eff_love}), we still find significant biases in the inferred tidal deformability, emphasizing the importance of including nonlinear hydrodynamics for accurate parameter estimation.

Our formulation of Newtonian equilibrium and dynamical tides enables the study of high-frequency phenomena like mode resonance, where full GR treatment is not yet available. It can also be used to identify additional effective Love numbers for the approach where the adiabatic Love number is replaced with a frequency-dependent, effective Love number as a simple way of extending the domain of validity to higher frequency. The commonly used one, defined in Eq. (\ref{eq:kappa_2m}), preserves only the radial tidal interaction. To generate the full waveform using the energy balancing approach, two additional effective Love numbers should be introduced. One is for the total equilibrium energy of the system, which would include the mode energy term that involves a Lorentzian squared. The other is to account for the FF correction to the energy radiation, where only the $|m|=2$ NS quadrupole moments contribute to the energy lost due to NS-orbit coupling (eq. 146 of \cite{Yu:25a}). 

We note for future investigation that the TRC due to background spin enables tests of GR via the I-Love universal relation. This is possible, because the TRC probes the spin frequency, $\Omega_\mathrm{A}$, while the PP-orbit waveform depends on the spin angular momentum, $S_\mathrm{A}=I_\mathrm{A}\Omega_\mathrm{A}$. Their ratio yields a direct measurement of $I_\mathrm{A}$, which can be used to test the I-Love relation.

Beyond testing GR, the background spin TRC offers an independent inference of $I$ solely from GWs. Such measurements of $I$ provide cross-validation of X-ray observations, \cite{Watts:16, Miller:19, Riley:19, Riley:21, Raaijmakers:21}.

\ack{The authors acknowledge Neil Cornish for insightful discussion and Aiden Gundersen for contributing initial code. We also thank the anonymous referees for their insightful and helpful feedback, which helped improve the quality and clarity of this manuscript.}

\funding{This work is supported by NSF grant No. PHY-2308415, PHY-2541579, and Montana NASA EPSCoR Research Infrastructure Development under award No. 80NSSC22M0042.}


\data{The code used to produce the results presented in this paper is available upon reasonable request.}


\bibliographystyle{plain}
\bibliography{ref}

\appendix

\section{Tidal Phase Shift Derivation}
\label{app:tidal_phase_shift_derivation}

We calculate the frequency-domain GW waveform phase shift due to tidal effects by using an energy balancing approach outlined in \S\ref{sec:tidal_phase_shift}, culminating in Eqs. (\ref{eq:delta_tc}, \ref{eq:delta_phic}). This approach requires calculating the orbital energy, GW radiation, and corresponding tidal corrections which are numerically integrated over GW frequency to arrive at the tidal phase shift. For quantities that vary during the inspiral, we calculate them along a GW frequency vector, from 100 Hz to 2 kHz.

We start with defining the orbital energy for a quasi-circular, Keplerian orbit, treating the NSs as PPs,
\begin{equation}\label{eq:E_orb}
    E_\mathrm{orb} =-\frac{1}{2}\frac{M_\mathrm{A} M_\mathrm{B}}{r} \,.
\end{equation}
We use the PP orbital frequency evolution due to GW radiation following eqs. (66, 67) in \cite{Yu:24a},
\begin{eqnarray} \label{eq:omega_dot_pp}
    \dot{\omega}_{\mathrm{pp}}&=&\frac{96}{5} \frac{M_\mathrm{A} M_\mathrm{B}}{M_\mathrm{tot}^{4}}\left(M_\mathrm{tot} \omega\right)^{11 / 3} \,, \\
    \label{eq:r_dot_pp}
    \frac{\dot{r}_{\rm{pp}}}{r}&=&-\frac{2}{3}\, \frac{\dot{\omega}_{\rm{pp}}}{\omega} \;=-\frac{\dot{E}_{\rm{pp}}}{E_{\rm{orb}}} \,,
\end{eqnarray}
where the GW luminosity of the PP orbit, $\dot{E}_{\rm{pp}}$, follows the same frequency evolution.
The orbital energy and GW luminosity provide the PP component to the orbital evolution but miss the finite-size effects which manifest as tides.

Next we consider the finite size of the NSs which results in corrections to the conserved energy of the BNS system. These tidal corrections to the energy, given by Eq. (\ref{eq:Delta_E}), $\Delta E = \Delta E_{\mathrm{orb}} + E_{\mathrm{int}} + E_{\mathrm{mode}}$, include a correction to the orbital energy, the tidal interaction energy, and the tidal mode energy. Tides are introduced via an interaction energy that couples the tidal field of the companion NS to the tidal amplitude in the primary NS. This interaction pulls energy out of the orbit and into tidal mode amplitude, resulting in a back-reaction on the orbit which modifies the orbital energy.

To quantify the tidal back-reaction on the orbit, we first calculate the radial acceleration imparted on the orbit by the tides pulling energy out of the orbit via the interaction Hamiltonian following eq. (28) of \cite{Yu:23a},
\begin{equation} \label{eq:gr_tide}
    g_{r}^\mathrm{(tide)}\simeq-\frac{E_\mathrm{A}}{\mu r} \sum_{m}^{\{2,0,-2\}}\,\sum_a^{m_a=m}6 \operatorname{Re}\!\left[V_{a}^*\, b_{a}^\mathrm{(eq)}\right]\,.
\end{equation}
The internal energy ($E_\mathrm{A} = M_\mathrm{A}^2/R$) for each NS is denoted by $\mathrm{A}$ or $\mathrm{B}$. The equilibrium tide amplitude, $ b_{a}^\mathrm{(eq)}$, is given by Eq. (\ref{eq:eq_tide}). We drop the dynamical tide component, because it is small (as we focus on non-resonant f-modes) and has a time dependence that is not phase-coherent with the orbit. Dropping the dynamical tide here means we ignore tidally induced eccentricities to the orbit. The time dependence of the equilibrium tide and tidal potential cancel out. The tidal potential for each NS becomes
\begin{equation}\label{eq:V_a,2m}
    V_{a} = W_{2m} I_{a} \frac{M_\mathrm{B}}{M_\mathrm{A}} \left(\frac{R}{r}\right)^3 \,,
\end{equation}
where $W_{2m}=\{\sqrt{\frac{3\pi}{10}},-\sqrt{\frac{\pi}{5}},\sqrt{\frac{3\pi}{10}}\}$ for $m=\{2,0,-2\}$, and $I_{a}=\sqrt{k_{2} \frac{5}{4\pi}}$. We replace the adiabatic quadrupolar Love number with tidal deformability, $k_{2}=\frac{3}{2}\bar{\lambda}_{2}(M_\mathrm{A}/R)^{5}$. The NS separation can be replaced with orbital frequency via Kepler's third law to linear order in $\Delta r/r$.

We recast the radial acceleration on the orbit due to tidal back-reaction as a correction to the NS separation, following eq. (39) in \cite{Yu:23a},
\begin{equation}\label{eq:delta_r_gr}
    \frac{\Delta r}{r} \simeq-\frac{1}{3} \frac{r^{2} g_{r}^\mathrm{(tide)}}{M_{\mathrm{tot}}} \,.
\end{equation}
This correction to the Kepler orbit is positive ( $g_r^{\mathrm{(tide)}}<0$ ), meaning the NSs must be farther apart to account for the additional inward acceleration while maintaining the same orbital frequency.

There are two tidal corrections that are proportional to this modified NS separation: the orbital energy correction $\Delta E_\mathrm{orb}$, see eq. (48) in \cite{Yu:23a}, and the interaction energy $E_\mathrm{int}$, see eq. (49) in \cite{Yu:23a},
\begin{equation}
    \Delta E_{\mathrm{orb}} = -4 \frac{\Delta r}{r} E_\mathrm{orb}\,, \qquad E_{\mathrm{int}} = 2 \frac{\Delta r}{r} E_\mathrm{orb}\,.
\end{equation}

The last tidal correction to the energy is the tidal mode energy. We use eq. (43) of \cite{Yu:23a} but corrected for spin as shown in eq. (58) of \cite{Yu:24a},
\begin{equation}
    E_{\mathrm{mode}}= E_\mathrm{A}\sum_{m_a}^{\{2,0,-2\}} \frac{\omega_{a\mathrm{S}}+m_{a} \Omega_\mathrm{A}}{\omega_{a0}} \left|b_{a}\right|^{2}\,.
\end{equation}
Here we use the full tidal amplitude, equilibrium and dynamical tide, denoted by $b_a$, which is calculated later in this section. As the tidal forcing frequency approaches mode resonance, $E_{\rm mode}$ is mainly driven by a tidal torque (eqs. 62-64 of \cite{Yu:24a}), so accounting for $E_{\rm mode}$ in the total energy allows us to bypass the calculation of the tidal torque in the equation of motion.

Next we consider the dissipative tidal corrections to the GW radiation given by Eq. (\ref{eq:Delta_E_dot}), $\Delta \dot{E}=\Delta \dot{E}_{\mathrm{ns-orb}}+\Delta \dot{E}_{r-\omega}$.
The first correction is the coupling between the tidally induced NS quadrupole and the orbital quadrupole, and the second correction accounts for the modified NS separation of Eq. (\ref{eq:delta_r_gr}). These two are the dominant corrections, and both enhance GW radiation, resulting in a shorter orbital evolution.

We use eq. (56) of \cite{Yu:23a} for the NS-orbit coupling correction,
\begin{equation}
    \Delta \dot{E}_{\mathrm{ns-orb}} \simeq -\frac{8}{15} \eta \frac{M_\mathrm{A} R^2}{M_{\mathrm{tot}}^3} (M_{\mathrm{tot}} \omega)^{14/3}\, \sqrt{\frac{3\pi}{10}} I_{a} \sum_{m_a}^{\{2,-2\}} m_a^6\operatorname{Re}\! \left[ b_{a}^{(\mathrm{eq})} \right]\,.
\end{equation}
We again only use the equilibrium tide here for the same reason in Eq. (\ref{eq:gr_tide}). Note the sum over $m_a$ does not include the $m_a=0$ term due to the expression being proportional to $m_a^6\,$; that mode does not radiate GWs.

For the last dissipative correction due to the modified orbit, we use eq. (59) in \cite{Yu:23a},
\begin{equation}
    \Delta \dot{E}_{r-\omega}=4 \frac{\Delta r}{r} \dot{E}_{\mathrm{pp}}\,.
\end{equation}

The total energy ($\Delta E$) and GW radiation ($\Delta \dot{E}$)  corrections can now be assembled for each NS.
Eqs. (\ref{eq:E_orb}, \ref{eq:r_dot_pp}, \ref{eq:Delta_E}, \ref{eq:Delta_E_dot}) are then combined and numerically integrated over GW frequency using Eqs. (\ref{eq:delta_tc}, \ref{eq:delta_phic}) to get the contribution to the GW waveform phase for each NS,
\begin{equation}
    \Psi_\mathrm{tide,A} = 2\pi f_\mathrm{gw} \Delta t_\mathrm{A} - \Delta \varphi_\mathrm{A} \,.
\end{equation}

The tidal contribution of the companion NS follows the same steps with the transformation $\mathrm{A}\leftrightarrow \mathrm{B}$. The total tidal contribution to the GW waveform phase is the sum of both NS contributions,
\begin{equation}
    \Psi_\mathrm{tide} = \Psi_\mathrm{tide,A} + \Psi_\mathrm{tide,B} \,,
\end{equation}
which can be plugged into Eq. (\ref{eq:waveform-phase}) for calculating the waveform phase.

What remains is to calculate the equilibrium and dynamical tide amplitudes which is where the TRCs enter the waveform. The equilibrium tide is defined in Eq. (\ref{eq:eq_tide}). We use the PP orbit evolution given by Eqs. (\ref{eq:omega_dot_pp}) and (\ref{eq:r_dot_pp}) for $\dot{\omega}$ and $\dot{r}$. 

Next is the dynamical tide amplitude, $c_{a}^{(\mathrm{dyn})}$, given by Eq. (\ref{eq:dyn_tide}). The dynamical tide requires evaluating quantities at resonance, such as Eq. (\ref{eq:omega_dot_pp}), 
\begin{equation} \label{eq:omega_dot_pp_r}
    \dot{\omega}_{\mathrm{pp,r}}\simeq\,\dot{\omega}_{\mathrm{pp}}(\omega=\omega_\mathrm{r}^{(1)}) \,.
\end{equation}
where resonance is denoted by the subscript ``$\mathrm{r}$", and we use $\omega_\mathrm{r}^{(1)}$ from Eq. (\ref{eq:omega_r^1}) to approximate the resonance frequency. We also evaluate the tidal potential at resonance,
\begin{equation}\label{eq:V_a,2m,r}
    V_{a,\mathrm{r}} \simeq V_{a}(r=r_\mathrm{r})\,, \quad \text{where} \quad r_\mathrm{r}=\left(\frac{M_\mathrm{tot}}{(\omega_\mathrm{r}^{(1)})^2}\right)^{1/3}  \,,
\end{equation}
via Kepler's third law. Note that corrections to $V_a$ due to GR and nonlinear tide are ignored as they are less important than the frequency shifts.

The numerical integral in $c_{a}^{(\mathrm{dyn})}$ is defined in eq. (24) of \cite{Yu:24a} as
\begin{equation} \label{eq:Fu}
    F(u)\coloneq\int_{-\infty}^{u} \frac{2 i\left(1+a_{1} u\right)}{\left(1+a_{2} u+2 i u^{2}\right)^{2}} e^{-i u^{2}} d u \,,
\end{equation}
where the constants $a_1$ and $a_2$ are defined below. Our choice of orbit direction leads to only the $m_a=2$ mode contributing to the dynamical tide, hence we simplify with $s_a=\operatorname{Sign}[m_a]=1$. The integration variable is a scaled time offset by the resonance time, following eqs. (25, 26) of \cite{Yu:24a}, 
\begin{equation} \label{eq:u}
    u\coloneq\left(\sqrt{\dot{\omega}_{\mathrm{pp}, \mathrm{r}}}\right) \tau\,,
\end{equation}
where 
\begin{equation}
    \tau\coloneq t-t_\mathrm{r}\simeq\frac{5}{256} \frac{M_\mathrm{A} M_\mathrm{B}}{M_\mathrm{tot}}\left[\left(M_\mathrm{tot} \omega_\mathrm{r}^{(1)}\right)^{-8 / 3}-\left(M_\mathrm{tot} \omega\right)^{-8 / 3}\right] .
\end{equation}

The constants reduce to,
\begin{equation}
    a_{1}=-\frac{5}{6}\sqrt{\dot{\omega}_\mathrm{pp,r} /\left(\omega_\mathrm{r}^{(1)}\right)^2} \,, \qquad a_2 = -2 \sqrt{\dot{\omega}_\mathrm{pp,r} /\left(\omega_\mathrm{r}^{(1)}\right)^2} \,.
\end{equation}

Finally we define a dynamical tide amplitude with the orbital phase factored out using eq. (30) in \cite{Yu:24a},
\begin{equation}\label{eq:b_a_dyn}
    b_{a}^{(\mathrm{dyn})} \simeq c_{a}^{(\mathrm{dyn})} e^{i[m_a(\phi-\phi_\mathrm{r})-\omega_\mathrm{r}^{(1)}\tau]}\,,
\end{equation}
where the difference between the instantaneous PP orbital phase and the orbital phase at resonance is given by eq. (31) of \cite{Yu:24a},
\begin{equation}
    \left(\phi-\phi_\mathrm{r}\right)_{\mathrm{pp}}=\frac{1}{32} \frac{M_\mathrm{tot}^2}{M_\mathrm{A} M_\mathrm{B}}\left[\left(M_\mathrm{tot} \omega_\mathrm{r}\right)^{-5 / 3}-(M_\mathrm{tot} \omega)^{-5 / 3}\right] \,.
\end{equation}
The dynamical tide in Eq. (\ref{eq:b_a_dyn}) is defined such that it can be directly added to the equilibrium tide to get the total tidal amplitude,
\begin{equation}\label{eq:b_a}
    b_a = b_{a}^{(\mathrm{eq})} + b_{a}^{(\mathrm{dyn})} \,.
\end{equation}

Now the tidal amplitudes, $b_a$ and $b_a^\mathrm{(eq)}$, can be calculated for each NS and used to find the tidal phase shift.

\section{Numerical Effective Love Number}
\label{app:eff_love_numerical}

We determine the numerical effective Love number by evolving the orbital equations of motion, Eqs. (\ref{eq:eom_r_ddot}, \ref{eq:eom_phi_ddot}), alongside the governing equations for the mode amplitudes, Eq. (\ref{eq:mode_gov_eq}). The orbital dynamics include non-Keplerian accelerations, for which we adopt the Burke-Thorne terms following eqs. (C21, C22) of \cite{Yu:23a}: 
\begin{eqnarray}
    g_r &=& 0 \,, \\
    g_\phi &=& -\frac{32}{5} \mu r^3 \dot{\phi}^5 \,.
\end{eqnarray}

We evolve the $\{l,m\}=\{2,2\}$ and $\{l,m\}=\{2,-2\}$ f-modes, whose amplitudes are denoted as $b_{2}$ and $b_{\text{-}2}$, respectively. They are needed to compute $k_2(2\omega)/k_2$ following the notation of Pitre and Poisson \cite{Pitre:25}, the effective Love number evaluated at twice the orbital frequency. The $m=0$ f-mode enters the nonlinear corrections of the $m=\pm 2$ modes, and we reduce them by their linear solution, which ensures our calculation is truncated at the three-wave order so that fair comparisons with Pitre and Poisson can be made. From Yu et al. 2023 \cite{Yu:23a}, the equations of motion of them read:
\begin{align}\label{eq:b2_dot}
    \dot{b}_2 &= -i(\omega_{a0} - 2\dot{\phi})b_2 + i\omega_{a0} \left[ W_{22} I_a \epsilon + W_{20} (J_2 + 4\kappa_2 I_a) \epsilon (b_2 + b_{\text{-}2}^\ast) \delta_{\rm nl} - \frac{\sqrt{6}\pi}{5} I_a J_2 \epsilon^2 \delta_{\rm nl} \right], \\
    \label{eq:b-2_dot}
    \dot{b}_{\text{-}2} &=-i(\omega_{a0} + 2\dot{\phi})b_{\text{-}2} + i\omega_{a0} \left[ W_{22} I_a \epsilon + W_{20} (J_2 + 4\kappa_2 I_a) \epsilon (b_{\text{-}2} + b_2^\ast) \delta_{\rm nl} - \frac{\sqrt{6}\pi}{5} I_a J_2 \epsilon^2 \delta_{\rm nl} \right].
\end{align}
Here, $\delta_{\rm nl}$ is a binary flag set to 1 to include nonlinear terms and 0 for the linear case. To simplify the notation, we define the parameter,
\begin{equation}
    \epsilon \coloneq \frac{M_{\rm B}}{M_{\rm A}} \left( \frac{R}{r} \right)^3 \,.
\end{equation}

The coupling coefficients adopted here are connected to Pitre and Poisson via 
\begin{align} \label{eq:k2_PP}
    k_2&= \frac{4\pi}{5} I_a^2,\\ 
    \label{eq:ddk2_PP}
    \ddot{k}_2&=\frac{4\pi}{5} I_a^2 \frac{M_{\rm A}/R_{\rm A}^3}{\omega_{a0}^2},\\ 
    \label{eq:p2_PP}
    p_2&=-\frac{8\pi}{5}\sqrt{\frac{\pi}{5}}(3J_2+4\kappa_2I_a),
\end{align}
with $\omega_{a0}^2/(M_{\rm A}/R_{\rm A}^3)=1.5$, $I_a=0.32$, $J_2=-0.21$, and $\kappa_2=-0.44$ (table 1 of Yu et al. 2023). This connection is established by matching the low-frequency expansion of the effective Love number; Eq. (\ref{eq:k22_lf_BY}). One can further check that the numerical values agree well.

The solutions to Eqs. (\ref{eq:b2_dot}, \ref{eq:b-2_dot}) are combined to calculate the quadrupolar moment of the primary NS,
\begin{equation}
    Q_{22} = (I_a + J_2 (c_0 + c_0^\ast))(b_2 + b_{\text{-}2}^\ast) \,,
\end{equation}
where $c_0 \coloneq W_{20} I_a \epsilon\,$. For the linear case, we evolve the  equations of motion with $\delta_{\rm nl}=0$, 
\begin{equation}
    Q_{22, \text{lin}} = I_a (b_{2, \text{lin}} + b_{\text{-2,lin}}^\ast)\,.
\end{equation}
Quantities with the ``${\rm lin}$'' subscript are calculated using the linear governing equations ($\delta_{\rm nl}=0$).

The numerical effective Love numbers, with and without nonlinear terms, are then obtained by normalizing the quadrupoles by the linear, adiabatic value,
\begin{equation}
    Q_{22, \text{adi,lin}} = I_a (2 W_{22} I_a \epsilon) \,,
\end{equation}
to get,
\begin{align}
    k_{\rm 22N,nl} &= Q_{22}/Q_{\rm 22,adi,lin} \,,\\
    k_{\rm 22N,lin} &= Q_{\rm 22,lin}/Q_{\rm 22,adi,lin} \,.
\end{align}
Numerical solutions of mode amplitudes are used to extract the values of $k_{\rm 22N,nl}$ (red-dashed line in Fig. \ref{fig:eff_love}) and $k_{\rm 22N,lin}$ (purple-dashed line). 

Eqs. (\ref{eq:b2_dot}) and (\ref{eq:b-2_dot}) can be solved analytically by noticing  $|\dot{b}_{\pm 2}| \ll (\omega_{a0}\mp2\dot{\phi})b_{\pm 2}$ before resonance, leading to an effective Love number that is corrected by the nonlinear tide in both denominator (frequency shift of the mode) and numerator (corrections to the tidal drive),
\begin{equation} \label{eq:keff_numer}
    k_{\rm 22BY} = \frac{1 + 4 W_{20} J_2 \frac{M_{\rm B}}{M_{\rm tot}} \frac{\omega^2}{\omega_{\rm A}^2}}{1 - 4\frac{\omega^2}{\omega_{a0}^2} - 2 W_{20} (J_2 + 4 \kappa_2 I_a) \frac{M_{\rm B}}{M_{\rm tot}} \frac{\omega^2}{\omega_{\rm A}^2} + i \frac{\omega\gamma_{\text{eff}}}{\omega_{a0}^2}} \,,
\end{equation}
where $\omega_{\rm A}^2 :=M_{\rm A}/R_{\rm A}^3\,$ is the dynamical frequency for the primary NS. We additionally include an effective damping term caused by orbital evolution (Lai 1994 \cite{Lai:94a}; Yu et al. 2024 \cite{Yu:24a}). The effective damping term is defined as,
\begin{equation}
    \frac{\omega \gamma_{\text{eff}}}{\omega_{a0}^2} := \frac{2\omega}{\omega_{a0}^2} \left( \frac{6\dot{r}}{r} - \frac{\dot{\omega}}{\omega} - \frac{16\omega^2}{\omega_{a0}^2 - 4\omega^2} \frac{\dot{\omega}}{\omega} \right) \,,
\end{equation}
recalling that $\dot{r}$ and $\dot{\omega}$ are related via Eq. (\ref{eq:r_dot_pp}).

For comparison, we also provide the effective Love number used to estimate detectability and parameter bias in this paper. Recall this version contains only the nonlinear TRC in the denominator,
\begin{equation}\label{eq:keff_denom}
    k_{\rm 22BY, den} = \frac{1}{1 - 4\frac{\omega^2}{\omega_{a0}^2} - 2 W_{20} (J_2 + 4 \kappa_2 I_a) \frac{M_{\rm B}}{M_{\rm tot}} \frac{\omega^2}{\omega_{\rm A}^2} + i \frac{\omega\gamma_{\text{eff}}}{\omega_{a0}^2}} \,.
\end{equation}
This effective Love number is identical to Eq. (\ref{eq:keff_numer}) except for the dropped nonlinear correction in the numerator. Recall the definition of the nonlinear TRC parameter, $\mathcal{A}_{\rm nl} = - W_{20}(J_2+4\kappa_2 I_a)M_{\rm B}/M_{\rm tot}(\omega_{\rm ref}/\omega_{a0})^2$, to see how it encapsulates the leading-order nonlinear correction to the f-mode frequency.

Finally, we restate the effective Love number from \cite{Pitre:25} [see their eq. (8.9)], modified here with an imaginary effective damping term,
\begin{equation}
    k_{\rm 22PP} = \frac{1}{1 - \frac{1}{k_2}\left(4 \ddot{k}_2 + p_2 \frac{M_{\rm B}}{ M_{\rm tot}}\right) \frac{\omega^2}{\omega_{\rm A}^2} + i \frac{\omega\gamma_{\text{eff}}}{\omega_{a0}^2}}\,.
\end{equation}
We calculate their tidal coefficients with Eqs (\ref{eq:k2_PP}-\ref{eq:p2_PP}). 

To assist direct comparison with \cite{Pitre:25}, we also provide the analytical low-frequency limit of our solution:  
\begin{equation} \label{eq:k22_lf_BY}
    k_{\rm 22BY,LF} = 1 + 4\frac{\omega^2}{\omega_{a0}^2} + 2W_{20}(3J_2+ 4\kappa_2I_a)\frac{M_{\rm B}}{M_{\rm A}+M_{\rm B}} \frac{\omega^2}{\omega_{\rm A}^2}\,,
\end{equation}
cf. eq. (8.8) of Pitre and Poisson \cite{Pitre:25}. 

The effective Love numbers derived here are presented in Fig. \ref{fig:eff_love}.

\end{document}